\documentclass[preprint2]{aastex}

\begin{document}

\def\la{\mathrel{\hbox{\rlap{\hbox{\lower4pt\hbox{$\sim$}}}\hbox{$<$}}}}
\def\ga{\mathrel{\hbox{\rlap{\hbox{\lower4pt\hbox{$\sim$}}}\hbox{$>$}}}}

\newdimen\digitwidth
\setbox0=\hbox{-.}
\digitwidth=\wd0
\catcode `@=\active
\def@{\kern\digitwidth}

\title{VLA Observations of a New Population of Blazars}

\author{Hermine Landt}
\affil{Harvard-Smithsonian Center for Astrophysics, 60 Garden Street, 
Cambridge, MA 02138.}

\author{Eric S. Perlman} 
\affil{Joint Center for Astrophysics, University of Maryland, 
1000 Hilltop Circle, Baltimore, MD 21250.}

\author{Paolo Padovani}
\affil{European Southern Observatory, Karl-Schwarzschild-Str. 2, 
D-85748 Garching, Germany.}

\begin{abstract}

We present the first deep VLA radio images of flat-spectrum radio
quasars (FSRQ) with multiwavelength emission properties similar to
those of BL Lacs with synchrotron X-rays. Our observations of
twenty-five of these sources show that their radio morphologies are
similar to those of other radio quasars. However, their range of
extended powers is more similar to that of BL Lacertae objects (BL
Lacs) and extends down to the low values typical of FR I radio
galaxies. Five out of our nine lobe-dominated sources have extended
radio powers in the range typical of both FR I and FR II radio
galaxies, but their extended radio structure is clearly FR
II-like. Therefore, we have not yet found a large population of radio
quasars hosted by FR Is. Two thirds of our sources have a
core-dominated radio morpholgy and thus X-rays likely dominated by the
jet. We find that their ratios of radio core to total X-ray luminosity
are low and in the regime indicative of synchrotron X-rays. This
result shows that also blazars with strong emission lines can produce
jets of high-energy synchrotron emission and undermines at least in
part the ``blazar sequence'' scenario which advocates that particle
Compton cooling by an external radiation field governs the frequency
of the synchrotron emission peak.

\end{abstract}


\section{Introduction}

Blazars are one of the most extreme classes of active galactic nuclei
(AGN). Their broad-band emission extends from radio up to gama-ray
frequencies, varies rapidly and irregularly, is strongly polarized,
and shows a core-dominated morphology. Furthermore, apparent
superluminal component speeds are often observed in these sources. The
properties of blazars can be best explained if we assume that, in
fact, they are classical double-lobed radio galaxies with their jets
seen at small angles relative to our line of sight and so subject to
strong relativistic beaming [see \citet{Urry95} for a review].

Based on their optical spectra, blazars are currently separated into
BL Lacertae objects (BL Lacs), defined to have no or only very weak
emission lines, and flat-spectrum radio quasars (FSRQ), defined to
have strong narrow and broad emission lines [but see \citet{L04} for a
suggested revision of the blazar classification scheme]. Within the
framework of unified schemes, BL Lacs are identified primarily with
beamed low-power Fanaroff-Riley type I \citep[FR I;][]{Fan74} radio
galaxies, whereas the high-luminosity Fanaroff-Riley type II (FR II)
radio galaxies are assumed to constitute the parent population of
radio quasars.

Traditionally, BL Lacs were discovered in radio and X-ray surveys, and
in recent years more efficiently through joint radio-X-ray selections,
whereas radio quasars had only been looked for in radio surveys. Thus,
it is not surprising that the synchrotron emission of BL Lac jets was
found to have its maximum within a wide range of frequencies, from
IR/optical to UV/soft-X-ray energies, whereas only radio quasars with
synchrotron emission maxima at relatively low energies, and thus
X-rays dominated by inverse Compton emission, were known. This, in
fact, led to the so-called ``blazar sequence'' which proposed that the
stronger the particle Compton cooling by an external radiation field
(such as the one produced by, e.g., the accretion disk or broad
emission line region, both stronger in quasars), the lower the
frequency of the synchrotron emission peak \citep{Sam96, Fos98,
Ghi98}. In other words, radio quasars with X-rays dominated by
synchrotron emission were not expected to exist.

Two recent surveys have drastically changed our picture of blazar jet
physics. About 10\% of the FSRQ discovered in the Deep X-ray Radio
Blazar Survey \citep[DXRBS;][]{Per98, L01} and $\sim 30\%$ of the ones
identified in the {\it ROSAT} All-Sky Survey (RASS) - Green Bank
\citep[RGB;][]{LM98, LM99} have been shown to have spectral energy
distributions (SEDs) and broad-band emission properties similar to
those of high-energy peaked BL Lacs \citep{P03}. Subsequent
investigations of the {\it EINSTEIN} Medium Sensitivity Survey (EMSS)
and the Slew survey proved that these ``X-ray-strong'' FSRQ had indeed
gone undetected in previous surveys \citep{Wol01b, Per01}.

In this paper we present the first deep radio interferometric
observations of this new class of blazars (Sections 2 and 3) and
discuss in the light of our observational results their nature and
their relation to ``standard'' FSRQ and BL Lacs (Section 4). Our
conclusions are presented in Section 5. For consistency with previous
work we have assumed throughout this paper cosmological parameters
$H_0 = 50$ km s$^{-1}$ Mpc$^{-1}$ and $q_0 = 0$. Spectral indices have
been defined as $S_\nu \propto \nu^{-\alpha}$.

\section{Observations and Data Reduction} \label{observe}

We have selected for observations with the NRAO\footnote{The National
Radio Astronomy Observatory is a facility of the National Science
Foundation operated under cooperative agreement by Associated
Universities, Inc.} Very Large Array (VLA) flat-spectrum ($\alpha_{\rm
r} \la 0.5$) radio quasars from the DXRBS and RGB surveys with
radio-to-X-ray spectral indices $\alpha_{\rm rx} \la 0.78$. The latter
constraint makes it likely that our sources have their X-rays
dominated by synchrotron rather than inverse Compton emission
\citep{P96}. We have preferably chosen sources at redshifts $z \la 1$
in order to minimize the effects of surface brightness dimming
\citep{Per93} and to increase the probability of detecting large-scale
extended radio emission. 

We were granted observing time for 25 sources. Their general
properties and log of observations are listed in Table
\ref{general}. The columns are: (1) object name; (2) redshift; (3)
total radio flux at 1.4 GHz from the NRAO/VLA Sky Survey (NVSS); (4)
radio spectral index between 1.4 and 5 GHz, calculated from the sum of
the fluxes of all NVSS sources within a $3'$ radius (corresponding
roughly to the beam size of the GB6 survey) and the total flux from
the GB6 and PMN surveys for northern and southern sources,
respectively; (5) unabsorbed {\it ROSAT} X-ray flux at 1 keV,
calculated using an X-ray spectral index derived from hardness ratios;
(6) $k$-corrected radio-to-X-ray spectral index between 5 GHz and 1
keV; (7) observation date; (8) VLA configuration; and (9) program
number. The observed sample contains 19/85 (or 22\%) low-redshift ($z
\la 1$) FSRQ with $\alpha_{\rm rx} \la 0.78$ from the combined RGB and
DXRBS samples. We expect the properties of the observed sources to be
representative of this entire subsample, since their total radio and
X-ray luminosities span similar ranges.

We observed with the VLA at 1.425 GHz in continuum mode. We imaged all
sources with the VLA in A configuration, and were able to gain C array
images for 14 of the 25 sources. Three or four scans of several
minutes length yielding a total exposure time of $\sim 45$ minutes
were interleaved with one minute scans on a suitable secondary VLA
flux calibrator for each source. Scans were spaced to optimize
coverage in the $(u,v)$ plane. Multiple observations of 3C 48, 3C 147,
or 3C 286 were used to flux-calibrate the maps.

The data were processed with the Astronomical Image Processing System
(AIPS; version 31DEC03) package. A model based on Clean components was
used to start the self-calibration process. Phase-only
self-calibration was used for the first three iterations and amplitude
and phase self-calibration for the last one or two iterations. The
task IMAGR with robust weighting (\mbox{ROBUST}=0.5) was used to
generate the maps and Clean components. We found it useful to combine
the self-calibrated data sets from the A and C arrays in order to
increase the sensitivity and to improve the $(u,v)$ plane coverage in
only 4/14 cases. Our results are shown in Figure 1 and the
corresponding map parameters are listed in Table \ref{map}, where the
columns are as follows: (1) object name; (2) VLA configuration; (3)
beam size; (4) position angle; (5) image rms; (6) peak flux; and (7)
corresponding panel in Figure 1. The dynamic ranges (peak/noise) of
the maps lie between $\sim 100-5000$.

The high spatial resolution provided by the A array data allowed a
good measure of the core emission, and thus in combination with the C
array data an accurate determination of the extended emission flux
density, morphology and size. In Table \ref{radioprop} we list the
radio properties of the sources as derived from our observations. The
columns are: (1) object name; (2) and (3) position of the unresolved
core, measured on A array map before start of self-calibration
process; (4) core flux density, measured on final A array map; (5)
extended flux density, derived from the total flux density measured on
C array map (where available and if the source was resolved, else
measured on A array map) substracting the core flux density given in
column (4); (6) core luminosity $k$-corrected with spectral index
$\alpha_{\rm r}=0$; (7) extended luminosity $k$-corrected with
spectral index $\alpha_{\rm r}=0.8$; (8) radio core dominance
parameter $R$, defined as $R=L_{\rm core}/L_{\rm ext}$, where $L_{\rm
core}$ and $L_{\rm ext}$ are the core and extended luminosities,
respectively; (9) ratio of NVSS flux to total flux density measured on
C array map (where available, else measured on A array map); and (10)
largest angular extent (LAS), measured from the core to the peak in
the extended radio structure [following \citet{Mur93}] on C array map
(where available, else on A array map). We give $1\sigma$ errors and
$2\sigma$ upper limits. The errors in the core flux densities are on
average $\sim 0.07$ mJy, close to the expected thermal noise. The
cumulative errors in the extended flux densities depend upon the solid
angular extent of the measured flux. For a considerable number of our
sources (11/25 objects) we have only A array data available, which
might miss extended flux. We have assessed this effect by using the
distribution of values in column (6), i.e., the ratio of NVSS flux to
total flux density measured. This has a median of 1.02 and
$\sigma=0.21$. Based on this we have assumed the NVSS flux as the
correct total flux for sources with ratios $\ga 1.5$ (the median value
plus $2 \sigma$). This applies to only one source, namely
WGAJ0435$-$0811, for which we list in column (5) the 'corrected'
extended flux.

\section{Discussion of Individual Sources}

We give now for the individual sources a brief discussion of their
general properties and a description of the radio images.

\vspace*{0.2cm}

{\sl RGB J0112$+$3818. $-$} An optical spectrum of this source was
published in \citet{LM98} based on which it was classified as a
quasar. This source is also part of the CLASS blazar survey
\citep{Marcha01, Cac02}. Faint extended radio emission is visible in
both images, and most pronounced in the A array image which shows two
faint lobes close to the core (Fig. 1(a)).

\vspace*{0.1cm}

{\sl RGB J0141$+$3923. $-$} Spectroscopic observations were published
in \citet{Marz03}. This source is also part of the CLASS blazar survey
\citep{Marcha01, Cac02}. The source remains unresolved in the radio on
the scales imaged (Fig. 1(b)).

\vspace*{0.1cm}

{\sl RGB J0254$+$3931. $-$} Optical spectra can be found in
\citet{Marcha96} and \citet{LM98} who classified this source as a
broad-line radio galaxy and quasar respectively. This source shows a
one-sided jet on the combined A$+$C image which is most prominent on
the A array image and appears to end in a lobe or hot spot (Figs. 1
(c) and (d)). A similar morphology is visible also on the map
published by \citet{Tay96} based on VLA B array snapshot observations
at 4.535 GHz.

\vspace*{0.1cm}

{\sl RGB J1629$+$4008. $-$} An optical spectrum of this source was
published in \citet{LM98} who classified it as a broad-line radio
galaxy based on an absolute optical magnitude of $M>-22$ mag. X-ray
observations of this source were obtained with {\sl Beppo}SAX by
\citet{P02}. The source remains unresolved in the radio on the scales
imaged (Fig. 1(e)). Similarly, the VLA B array map from the FIRST
survey at 20 cm shows only an unresolved core.

\vspace*{0.1cm}

{\sl RGB J2229$+$3057. $-$} \citet{LM98} presented an optical spectrum
of this source and classified it as a quasar. The source is extended
on both A and C array images (Figs. 1(f) and (g)). In particular, on
the A array image two well-defined lobes are visible at each side of
the nucleus, the closest lobe being fairly extended with a hot spot.

\vspace*{0.1cm}

{\sl RGB J2256$+$2618. $-$} An optical spectrum of this source was
published by \citet{LM98} who classified it as a broad-line radio
galaxy based on an absolute optical magnitude of $M>-22$ mag.
Extended emission on scales of several tens of arcseconds is detected
in the C array image (Fig. 1(h)).

\vspace*{0.1cm}

{\sl RGB J2308$+$2008. $-$} An optical spectrum can be found in
\citet{LM98}. Based on this and an absolute optical magnitude of
$M>-22$ mag these authors classified this source as a broad-line radio
galaxy. \citet{Falco98} give a slightly different redshift for this
source of $z=0.2342$ based on four emission lines. The source is
unresolved on C array scales, however, a faint lobe is visible close
to the core on the A array image (Fig. 1(i)).

\vspace*{0.1cm}

{\sl RGB J2318$+$3048. $-$} \citet{LM98} published an optical spectrum
of this source and classified it as a broad-line radio galaxy based on
an absolute optical magnitude of $M>-22$ mag. This source is clearly
extended on the C array image (Fig. 1(j)), however, only the core is
detected on A array scales.

\vspace*{0.1cm}

{\sl WGA J0106$-$1034. $-$} Optical spectra of this source can be
found in \citet{Wol98} and \citet{L01} who classify it as a
broad-emission line AGN and quasar, respectively. This source is also
part of the FIRST Bright Quasar Survey \citep{Beck01} and the Sloan
Digital Sky Survey \citep{Schn03}. The A array image of this source
(Fig. 1(k)) shows extended emission with hot spots on each side of the
nucleus, making it clearly an FR II source. However, one of the two
hot spots is considerably stronger than the other, and this source
appears unusual in that the radio jet we see points towards the weaker
of the two lobes. The source is unresolved on C array scales. Our ATCA
snapshot observations of this source give a steeper spectral index of
$\alpha_{\rm r}=0.67$ (between 4.8 and 8.6 GHz) than the one listed in
Table \ref{general} ($\alpha_{\rm r}=0.35$).

\vspace*{0.1cm}

{\sl WGA J0110$-$1647. $-$} Optical spectra of this source can be
found in \citet{Per98} and \citet{Cac00} who classify it as a quasar
and broad-line object, respectively. An extended jet-like feature is
clearly visible on the A array image (Fig. 1(l)) a few arcseconds away
from the core.

\vspace*{0.1cm}

{\sl WGA J0126$-$0500. $-$} An optical spectrum of this source was
published by \citet{L01} who classify it as a quasar. The A array
image (Fig. 1(m)) shows it to be a double-lobed source with some
extended emission also around the core.

\vspace*{0.1cm}

{\sl WGA J0227$-$0847. $-$} An optical spectrum of this source can be
found in \citet{L01} who classify it as a quasar. This source is also
part of the Sloan Digital Sky Survey \citep{Schn03}. Our ATCA snapshot
observations of this source give a spectral index of $\alpha_{\rm
r}=0.37$ between 4.8 and 8.6 GHz, which is steeper than the one listed
in Table \ref{general} ($\alpha_{\rm r}=-0.34$). This is either due to
variability or to the fact that this source is a Gigahertz
peaked-spectrum (GPS) candidate \citep[e.g.,][]{ODea98}. The source
remains unresolved in the radio on the scales imaged
(Fig. 1(n)). Similarly, the VLA B array map from the FIRST survey at
20 cm shows only an unresolved core.

\vspace*{0.1cm}

{\sl WGA J0259$+$1926. $-$} An optical spectrum of this source can be
found in \citet{Cac00} who classify it as a broad-line object. Our own
high-quality spectrum of this source will be published shortly. We
classify this source as a quasar. The A array image of this source
(Fig. 1(o)) shows a jet-like feature extending a few arcseconds away
from the core. A slightly extended morphology is visible even on the
scales imaged with the VLBA at 8.4 GHz \citep{Ojha04}.

\vspace*{0.1cm}

{\sl WGA J0304$+$0002. $-$} Optical spectra of this source were
published by \citet{Per98} and \citet{Cac00} who classified it as a
quasar and a broad-line object, respectively. This source is also part
of the Sloan Digital Sky Survey \citep{Schn03}. The source appears
extended on both the A and C array images (Fig. 1(p) and (q)) with two
well-defined lobes on each side of the nucleus and some fuzzy extended
emission on larger scales off one of the lobes. Our ATCA snapshot
observations of this source give a steeper spectral index of
$\alpha_{\rm r}=0.71$ (between 4.8 and 8.6 GHz) than the one listed in
Table \ref{general} ($\alpha_{\rm r}=0.40$).

\vspace*{0.1cm}

{\sl WGA J0435$-$0811. $-$} \citet{Per98} published an optical
spectrum and classified this source as a quasar. It is unresolved on A
array scales (Fig. 1(r)), however, most likely its extended emission
is resolved out. The NVSS image is slightly extended and the ratio of
NVSS to total measured flux on our image for this source is the
highest in the sample (see Table \ref{radio}, column (6)).

\vspace*{0.1cm}

{\sl WGA J0447$-$0322. $-$} Optical spectra of this source can be
found in \citet{Per98} and \citet{Cac00} who classified it as a quasar
and broad-line object, respectively. This source remains unresolved on
the scales imaged (Fig. 1(s)). Our ATCA snapshot observations give a
steeper spectral index of $\alpha_{\rm r}=0.91$ (between 4.8 and 8.6
GHz) than the one listed in Table \ref{general} ($\alpha_{\rm
r}=0.47$).

\vspace*{0.1cm}

{\sl WGA J0544$-$2241. $-$} \citet{Per98} published an optical
spectrum and classified this source as a quasar. The source remains
unresolved in the radio on the scales imaged (Fig. 1(t)). Our ATCA
snapshot observations give a spectral index of $\alpha_{\rm r}=0.31$
(between 4.8 and 8.6 GHz), which is steeper than the one listed in
Table \ref{general} ($\alpha_{\rm r}=-0.44$). This is either due to
variability or to the fact that this source is a Gigahertz
peaked-spectrum (GPS) candidate \citep[e.g.,][]{ODea98}.

\vspace*{0.1cm}

{\sl WGA J1026$+$6746. $-$} \citet{L01} published an optical spectrum
and classified this source as a quasar. The A array image (Fig. 1(u))
shows jet-like extended structure over several tens of arcseconds
emerging from the core to both sides and an extended lobe to the far
left.

\vspace*{0.1cm}

{\sl WGA J1457$-$2818. $-$} An optical spectrum of this source can be
found in \citet{L01} who classify it as a quasar. Slightly extended
emission away from the core is visible on A array scales (Fig. 1(v)).
In better accordance with the observed morphology, our ATCA snapshot
observations of this source give a steeper spectral index of
$\alpha_{\rm r}=0.72$ (between 4.8 and 8.6 GHz) than the one listed in
Table \ref{general} ($\alpha_{\rm r}=0.40$).

\vspace*{0.1cm}

{\sl WGA J2239$-$0631. $-$} \citet{L01} published an optical spectrum
and classified this source as a quasar. The source is extended on the
C array image (Fig. 1(w)) with two lobes to each side of the nucleus,
one of them having a stronger hot spot than the other. Only the
nucleus is detected on A array scales.

\vspace*{0.1cm}

{\sl WGA J2320$+$0032. $-$} An optical spectrum of this source can be
found in \citet{L01} who classify it as a quasar. This source is also
part of the Sloan Digital Sky Survey \citep{Schn03}. The source
remains unresolved in the radio on the scales imaged
(Fig. 1(x)). Similarly, the VLA B array map from the FIRST survey at
20 cm shows only an unresolved core.

\vspace*{0.1cm}

{\sl WGA J2322$+$2114. $-$} Optical spectra of this source can be
found in \citet{Wol98} and \citet{Per98} who classify it as a
broad-emission line AGN and quasar, respectively. The A array image
(Fig. 1(y)) shows clearly a double-lobed structure and a pronounced
core.

\vspace*{0.1cm}

{\sl WGA J2347$+$0852. $-$} \citet{Per98} published an optical
spectrum of this source and classified it as a quasar. This source is
extended in both our A and C array images (Figs. 1(z) and (aa)). The
X-shape of the galaxy is apparent only in the C array image, whereas
on A array scales we observe a central core with two almost
equidistant lobes. Both lobes have hot spots, however, one lobe is
brighter than the other.

\vspace*{0.1cm}

{\sl PKS 0256$-$005. $-$} The optical spectra of this source presented
by \citet{Rich80b} and \citet{Osm94} show several strong broad
emission lines. This source is also part of the FIRST Bright Quasar
Survey \citep{Beck01} and the Sloan Digital Sky Survey
\citep{Schn03}. The source remains unresolved in the radio on the
scales imaged (Fig. 1(ab)). Similarly, the VLA B array map from the
FIRST survey at 20 cm shows only an unresolved core.

\vspace*{0.1cm}

{\sl 0959$+$68W1. $-$} Optical spectra of this source are available
from the {\sl Hubble Space Telescope} archive. These show several
broad emission lines. This source is also part of the CLASS blazar
survey \citep{Marcha01, Cac02}. The A array map (Fig. 1(ac)) shows
jet-like features and jet blobs extending over several tens of
arcseconds on both sides of the pronounced core.

\section{Results and Discussion}

In this section we want to address three main questions: 1. How do the
general radio properties of these newly discovered strong-lined
blazars compare to those of ``classical'' radio quasars and BL Lacs?
2. Which type of radio galaxies form their parent population? and
3. Is the X-ray emission of these ``X-ray strong'' radio quasars
dominated by the jet and synchrotron in origin?

\subsection{The Comparison Samples} \label{samples}

For our studies we have selected for comparison other types of
radio-loud AGN, namely, steep-spectrum radio quasars (SSRQ),
``standard'' FSRQ and BL Lacs. 

The comparison sample of SSRQ has been drawn from the Molonglo Quasar
Sample (MQS) excluding compact steep-spectrum (CSS) quasars and
GHz-peaked sources (74 objects). \citet{Kap98} published for this
sample total and core radio fluxes, largest angular sizes, and radio
spectral indices (between 0.4 and 5 GHz) from observations with the
VLA in BnA and CnB configurations at 1.4, 5, and 8.4 GHz, and with the
Parkes-Tidbinbilla Interferometer (PTI) at 2.3 GHz. Largest angular
sizes were measured between the brightest peaks in the extended radio
structure. Improved {\sl ROSAT} X-ray fluxes were published by
\citet{Bak95b}.

The comparison sample of FSRQ has been selected from the list of
\citet{Mur93}. These authors published core and extended radio fluxes
and largest angular sizes for 56 sources from the 1 Jy catalogue
\citep{Kuehr81} as obtained from observations with the VLA in A and B
configurations at 1.4 GHz. The largest angular sizes were measured
from the core to the peak in the extended radio structure. We have
compiled X-ray fluxes at 1 keV and radio spectral indices between 1.4
and 5 GHz for these sources from the multiwavelength catalogue of
\citet{P97}.

Additionally, we have chosen for comparison the well-known 1 Jy and
EMSS BL Lac samples, since these have complete dedicated VLA
observations \citep[e.g.,][]{Per93,Cas99,Rec00,Rec01} as well as in
depth {\sl ROSAT} observations \citep[e.g.,][]{Per96a,Urry96,Rec99}
available. \citet{Per93} give for the EMSS sample largest linear sizes
defined as the sum of two straight lines from the core to the
outermost $3\sigma$ contours, with each line intersecting the
brightest hotspot, if any, in the extended structure. Simultaneous
radio spectral indices between 1.4 and 5 GHz for this sample have been
taken from \citet{Cava04}. \citet{Rec01} give for the 1 Jy sample
largest linear sizes defined as in \citet{Per93}. Radio spectral
indices between 2.7 and 5 GHz for this sample have been taken from
\citet{Sti93c}.

We have transformed core and extended radio fluxes to 1.4 GHz using a
radio spectral index of $\alpha_{\rm r} = 0$ and 0.8,
respectively. These same indices have been used to $k$-correct the
luminosities. We have transformed band fluxes to an energy of 1 keV
and $k$-corrected the corresponding luminosities using an X-ray
spectral index of $\alpha_{\rm x} = 1.2$.

\subsection{The Radio Properties} \label{radio}

In this section we investigate the radio properties of the newly
discovered ``X-ray strong'' radio quasars and compare them to those of
``classical'' radio quasars and BL Lacs.

\subsubsection{Extended Radio Powers and the Radio Core Dominance Parameter}

In Fig. \ref{logRlext} we have plotted for our sources and the
comparison samples introduced in Section \ref{samples} the extended
radio power at 1.4 GHz versus the radio core dominance parameter at
this frequency. The extended radio power is believed to be radiated
isotropically and, therefore, unaffected by relativistic beaming. Thus
this quantity is a good measure of the source's intrinsic jet
power. Also marked (horizontal solid lines) are the ranges of extended
power typical of FR I and FR II radio galaxies \citep{Owen94}. For a
constant extended radio power an increase in radio core dominance
parameter is indicative of stronger relativistic beaming, since
beaming affects only the core emission. For this reason the radio core
dominance parameter $R$ is assumed to be a suitable orientation
indicator (the higher $R$, the smaller the viewing angle, i.e., the
angle between jet and observer's line of sight).

Fig. \ref{logRlext} shows that, as expected, ``classical'' radio
quasars (SSRQ and FSRQ, filled triangles and squares, respectively)
have FR II-like extended powers, with only one FSRQ from the sample of
\citet{Mur93} having an extended power more typical of an FR I. On the
other hand, the BL Lacs from the samples considered here span a range
in extended power larger than that of quasars, populating both the FR
I- and FR II-regimes. Our sources differ from the radio quasars
plotted here. Their range of intrinsic radio power is similar to that
of BL Lacs and extends down to the values typical of FR Is, reaching
about three orders of magnitude lower values than ``classical'' radio
quasars.

The low radio powers of our sources are partly due to the fact that
they are selected from surveys with factors of $\sim20-80$ lower radio
flux limits than the comparison radio quasars. This selection,
however, gives us a most important result. We have found either the
lowest luminosity examples of FR II radio galaxies or quasars hosted
by FR Is. The first possibility would indicate that morphology and
intrinsic radio power of radio galaxies are not as tightly connected
as first found by \citet{Fan74}. The second possibility would mean
that FR I radio galaxies can host quasar nuclei, i.e., they can have
broad emission line regions at their centers. To our knowledge, so far
only three such objects are known, J2114$+$820, a broad-line radio
galaxy \citep{Lara99}, E1821$+$643, a quasar associated with a
subarcsecond jet in an FR I radio structure \citep{Blu01}, and the
well-known FR I broad-line radio galaxy 3C 120
\citep[e.g.,][]{Walker87}. In any case, our finding of a large number
of radio quasars with strong emission lines but relatively low radio
powers can be regarded as another observational fact against a close
relationship between emission line luminosity and jet power in AGN
\citep{Raw91}, in addition to the existence of radio-quiet quasars
(i.e., sources with strong emission lines but extremely weak radio
jets) and BL Lacs (i.e., sources with very weak emission lines but
strong radio jets).

Fig. \ref{logRlext} also shows that our sources with extended powers
in the range populated by the comparison radio quasars have $R$ values
intermediate between those of SSRQ and FSRQ. This result, however,
does not mean that ``X-ray strong'' quasars {\it are} the less beamed
versions of ``classical'' FSRQ. For a correct interpretation we need
to take the above mentioned selection effects into account. FSRQ
selected in surveys with lower radio flux limits (such as our sources)
will have on average lower {\it total} radio luminosities than FSRQ
selected in surveys with higher radio flux limits. Then, considering
only a narrow range, i.e., an almost constant value, of relatively
high {\it extended} radio powers, the former will have on average
lower {\it core} luminosities and so will appear less beamed than the
latter. In fact, the sources in our sample reach higher $R$ values
than the FSRQ in the comparison sample, but at lower extended radio
powers.

\subsubsection{Largest Linear Sizes}

In Fig. \ref{logRLLS} we have plotted for our sources and the
comparison samples introduced in Section \ref{samples} the largest
linear size (LLS) versus the radio core dominance parameter $R$ at 1.4
GHz. We note that the largest linear sizes were measured differently
in each sample. In the FSRQ and our sources they were measured from
the core to the peak in the extended structure, whereas in the SSRQ
and BL Lac samples they were measured between the peaks of the
extended structure on either side of the nucleus. However, we did not
attempt to correct the values for the SSRQ and BL Lacs since, on the
one hand, such a correction is not straightforward (e.g., assumptions
about the intrinsic radio structure are required) and, on the other
hand, we are interested mainly in comparisons between our sources and
the FSRQ.

As expected from unified schemes, Fig. \ref{logRLLS} shows that the
largest linear sizes of radio-loud AGN decrease with increasing radio
core dominance parameter, i.e., with decreasing angle between jet and
observer's line of sight. Our sources follow this trend very
well. Significant LLS $- R$ correlations have been reported previously
by other authors for radio quasars, but not for BL Lacs
\citep[e.g.,][]{Hine80, Kap82, Hough89, Hough99}. Following their
example we plot in Fig. \ref{logRLLS} also the relation between LLS
and $R$ expected from relativistic beaming models (dashed, dotted and
solid lines). In view of the selection effects and incompleteness in
the data we restrict ourselves to a prediction of the expected upper
and lower envelopes and do not attempt to extract an average or median
value of LLS as a function of $R$ to test beaming models.

The fits were obtained as follows. If $\theta$ is the angle between
the jet axis and the line of sight, the observed values of LLS and $R$
are given by, ${\rm LLS} = L_0 \sin \theta$ and $R = (R_{90}
B(\theta))/2$, where $L_0$ is the intrinsic largest linear size,
$R_{90}$ the radio core dominance parameter that would be observed if
the source was oriented at $\theta=90^{\circ}$, and $B(\theta) = (1 -
\beta \cos\theta)^{-(2+\alpha_{\rm r})} + (1 + \beta
\cos\theta)^{-(2+\alpha_{\rm r})}$ for the case of a continuous jet
with $\beta=v/c$, the ratio of jet velocity to the speed of light. The
relation between LLS and $R$ results from their common dependence on
viewing angle and its simulation requires initial values for $L_0$,
$R_{90}$, the Lorentz factor $\Gamma = 1/\sqrt{1-\beta^2}$, and the
radio spectral index $\alpha_{\rm r}$.

For the radio spectral index we have assumed in all cases a value of
$\alpha_{\rm r} = 0.8$, which is typical of optically thin synchrotron
emission. Radio galaxies can have different intrinsic radio core
dominance values and we have assumed values of $R_{90} = 0.022$ and
0.003, the median values obtained by \citet{Morg97} for their samples
of FR I and narrow line FR II radio galaxies, respectively. In the
case of $R_{90} = 0.003$, the upper and lower envelopes in
Fig. \ref{logRLLS} can be fitted if we assume a bulk Lorentz factor of
$\Gamma = 8$ and values of $L_0 = 1900$ and 30 kpc, respectively
(dashed lines). In the case of $R_{90} = 0.022$, a slightly lower bulk
Lorentz factor of $\Gamma = 6$ and values of $L_0 = 1400$ and 20 kpc,
respectively, are required (dotted lines). We note that a fit with a
lower radio spectral index would require a much higher value of the
bulk Lorentz factor, e.g., in the case of $\alpha_{\rm r} = 0.2$ and
$R_{90} = 0.003$ a value of $\Gamma = 15$.

Previously presented LLS $-$ $R$ relations included only
lobe-dominated radio quasars, i.e., sources viewed at relatively large
angles, but did not account for this by restricting the range of
orientation accordingly. Thus it is not suprising that fits to their
envelopes required smaller Lorentz factors of $\Gamma = 2-5$ (for
values of $R_{90}$ similar to our FR II radio galaxy case). We note
also that the sources in the samples considered here reach higher LLS
values than those included in the previous studies. We have included
in Fig. \ref{logRLLS} both quasars and BL Lacs and it appears that
they follow the same relation, but their starting points are most
probably different, which we have accounted for here by adopting two
cases of $R_{90}$. The resulting maximum intrinsic LLS values of their
parent radio galaxies are not very much different and in accordance
with the result of \citet{Led02} who found maximum sizes for both FR I
and FR II radio galaxies around 2000 kpc (with only one source
exceding this value).

%We now want to compare the average LLS values of our sources with
%those of ``classical'' FSRQ. For this purpose, we have excluded from
%our sample objects with only upper limits available (8 sources), since
%\citet{Mur93} did not include in their analysis objects with no
%detected extended emission. In this case, according to a
%Kolmogorov-Smirnov (KS) test, the mean largest linear size of ``X-ray
%strong'' radio quasars (17 objects) of $196 \pm 39$ kpc is
%significantly ($P=97.8\%$) higher than the mean LLS value of
%``classical'' FSRQ of $95 \pm 11$ kpc. However, our sources have also
%a significantly ($P>99.9\%$) lower mean radio core dominance parameter
%than the ``classical'' FSRQ ($\langle \log R \rangle = -0.04 \pm 0.17$
%and $1.15 \pm 0.10$, respectively). This most certainly accounts for
%the observed difference in mean LLS values: the subsample of our
%sources considered here is on average less beamed and so has on
%average larger sizes than the comparison FSRQ sample. This is
%illustrated in Fig. \ref{logRLLS}, where we plot the average values
%for our sources and the ``classical'' FSRQ as stars and as a solid
%line a fit based on the beaming model described above assuming the
%case of $R_{90} = 0.003$ and a source intrinsic size of $L_0 = 430$
%kpc.

\subsubsection{The Selection of Blazars}

Past and on-going blazar surveys, such as, e.g., the 1 Jy BL Lac
survey \citep{Sti91}, CLASS blazar survey \citep{Marcha01, Cac02} and
DXRBS \citep{Per98, L01}, have selected their candidates based on
radio spectral index. A limit of $\alpha_{\rm r} = 0.5$ is generally
chosen, since radio sources are expected to become increasingly
core-dominated below this value \citep{Orr82}. Ideally, blazar
selection should be based on orientation, but methods to derive actual
viewing angles require long-term radio observations and/or
sophisticated multiwavelength data sets. A blazar selection based on
the value of the radio core dominance parameter of $R>1$ would then be
another possibility, but this would require dedicated radio
observations for a large number of sources, which are usually not
readily available. Therefore, this approach would be extremely time
consuming.

In order to assess the effect of blazar selection based on radio
spectral index relative to one possibly based on the radio core
dominance parameter we have plotted in Fig. \ref{logRalr} for our
sources and the samples introduced in Section \ref{samples} these two
quantities versus each other. Fig. \ref{logRalr} shows that indeed
radio-loud AGN appear to populate preferably the regions of either
steep radio spectral index ($\alpha_{\rm r}>0.5$) and lobe-dominance
(i.e., $R<1$) or flat radio spectral index ($\alpha_{\rm r}<0.5$) and
core-dominance (i.e., $R>1$). Only a few sources in blazar surveys
imposing a radio spectral index cut of $\alpha_{\rm r} = 0.5$ are
expected to be lobe-dominated, and similarly, only a few
core-dominated sources are expected to be missed by these surveys.

A considerable fraction of our sources (7/25 or 28\%), however, have a
flat radio spectral index $\alpha_{\rm r}\le0.5$ and {\it are}
lobe-dominated. Are we then justified to refer to these sources as
blazars? Variability undoubtedly plays a role in the explanation of
the existence of at least some of the lobe-dominated flat-spectrum
radio sources, however, it is unlikely that it can account for the
existence of all of them. For example, we have additional ATCA
snapshot observations available for 3 out of the 7 sources, and for
two of them, namely, WGAJ0106$-$1034 and WGAJ0304$+$0002, these yield
steep radio spectral indices of $\alpha_{\rm r} = 0.67$ and 0.71,
respectively. For WGAJ0126$-$0500 our ATCA observations give
$\alpha_{\rm r} = 0.39$, only slightly steeper than the value of 0.21
listed in Table \ref{general}. A more plausible explanation for the
existence of lobe-dominated flat-spectrum radio sources can instead be
gained from simulations of the relation between $\alpha_{\rm r}$ and
$R$.

From the definitions of the radio spectral index between two
frequencies $\nu_1$ and $\nu_2$ of $(S_{\nu_1}/S_{\nu_2}) \propto
(\nu_1/\nu_2)^{-\alpha_{\rm r}}$ and the relation between the radio
core dominance parameters at these two frequencies of $R_{\nu_2} =
R_{\nu_1} (\nu_1/\nu_2)^{\alpha_{\rm r_c}-\alpha_{\rm r_e}}$, where
$\alpha_{\rm r_c}$ and $\alpha_{\rm r_e}$ are the radio spectral
indices of the core and extended radio emission, respectively, we have
derived $\alpha_{\rm r} - \log R_{\nu_1}$ relations for different
values of $\alpha_{\rm r_c}$. We have assumed for $\nu_1$ and $\nu_2$
values of 1.4 and 5 GHz, respectively, and for the extended emission a
radio spectral index of $\alpha_{\rm r_e}=0.8$. Our results are shown
in Fig. \ref{logRalr} as dashed lines from top to bottom for a core
radio spectral index of $\alpha_{\rm r_c}=0.5$, 0, $-0.5$, and $-1.3$.

Firstly, we note that the relation between $\alpha_{\rm r}$ and $\log
R$ is not a linear one \citep[see also][]{Kem86}, contrary to what the
fit of \citet{Cava04} to their Fig. 4 suggests. Secondly, our results
show that lobe-dominated flat-spectrum radio sources can indeed exist,
if the core emission has a radio spectral index of $\alpha_{\rm
r_c}\la0$. In general, the lower the core radio spectral index, the
less core flux relative to the extended emission is required to render
the total radio spectral index of the source below a value of
$\alpha_{\rm r_c}=0.5$. Therefore, a blazar selection and definition
based on a radio core dominance parameter value of $R\ga1$ appears to
be more robust than one based on a radio spectral index of
$\alpha_{\rm r}\la0.5$. But, as Fig. \ref{logRalr} shows, a selection
based on a flat radio spectral index is expected to include
lobe-dominated sources rather than to miss blazars.

In this regard, we note that the result that FR I radio galaxies have
higher $R$ values than FR II radio galaxies \citep[e.g.,][]{Morg97}
could be due to selection effects. If the cores of FR II radio
galaxies had on average lower radio spectral indices than the ones of
FR I radio galaxies, in samples selecting at relatively high
frequencies and thus sensitive to flat-spectrum sources, the first
will appear on average less core-dominated than the latter.

\subsection{The Parent Population} \label{parent}

According to current unified schemes for radio-loud AGN, weak-lined
blazars, the BL Lacs, are harboured mainly by low-power radio
galaxies, the FR Is, whereas radio quasars are hosted by
high-luminosity FR II radio galaxies. In Section \ref{radio} we have
shown that our sources extend the range of intrinsic radio powers of
quasars down to the low values typical of FR Is and BL Lacs. What then
is the parent population of our newly discovered ``X-ray strong''
radio quasars?

In our sample, 9/25 sources are lobe-dominated and in principle a
morphological classification of their extended radio structure is
possible. Out of these, 4 sources have FR II-like extended radio
powers with two of them, WGAJ0106$-$1034 and WGAJ2322$+$2114, having
also a clear FR II-like morphology. They show extended lobes with
prominent hot spots in their outer regions. The morphology of
WGAJ0304$+$0002 is not clear. One of the lobes appears to have an
outer hot spot and so is indicative of an FR II. The other lobe,
however, might be more FR I-like. A higher quality map of this object
would be helpful in this case. The source WGAJ1026$+$6746 shows an
extended jet structure with some extended lobe emission. We
tentatively interpret its morphology as FR II-like. Another 5 sources
have extended radio powers typical of both FR Is and FR IIs. Out of
these, RGBJ2229$+$3057, WGAJ0126$-$0500, and WGAJ2347$+$0852 show a
clear FR II-like morphology, especially on the A array map. The
extended structure of WGAJ2239$-$0631 appears to be more FR II-like,
however, only a B array image can assess this with certainty. The
source WGAJ0435$-$0811 needs to be imaged in B or C array in order to
classify its extended emission, since this is most likely resolved out
on the scales imaged (see Section \ref{observe}).

Unfortunately, all sources in our sample with extended powers typical
of FR Is (8/25 objects) are core-dominated and so a morphological
classification of their extended emission is not possible. Sources
with extended powers in this low regime and lobe-dominated morphology
are the most interesting ones, since they can clarify if a large
population of low-luminosity FR II radio galaxies and/or quasars
hosted by FR Is exist. In this respect, it is vital to observe with
the VLA more ``X-ray strong'' radio quasars and in general radio
quasars with low total radio powers in order to search for this kind
of objects.

%In this respect, a continuation of our observing program with the VLA
%of ``X-ray strong'' radio quasars is vital.

\subsection{The Origin of the X-ray Emission} \label{origin}

The X-ray emission of radio-loud AGN, contrary to their emission at
radio frequencies, is believed to be made up of at least three
components with different origins: an isotropic one, produced by the
ambient hot gas (presumably from the group or cluster associated with
the source or from the galaxy), and two anisotropic components, one
generally interpreted as inverse Compton emission from the accretion
disk corona, which is present also in radio-quiet quasars and is
expected to be seen only in sources with their nuclei unobscured by
the central dusty torus, and the other associated with the powerful
relativistic jet and thus subject to beaming.

In order to constrain the origin of the X-ray emission in our sources
we have plotted in Fig. \ref{lrclx} their ratio of radio core to total
X-ray luminosity versus the radio core dominance parameter $R$. A plot
such as the one displayed in Fig. \ref{lrclx} is most suitable for our
purpose. Relativistic beaming is believed to increase only the core
jet emission, both radio and X-ray, and leave any isotropically
emitted extended emission unaffected. If the total X-ray emission
$L_{\rm x}$ is dominated by an unbeamed component, the ratio $L_{\rm
core}/L_{\rm x}$ will then increase with $R$ until the beamed jet
emission starts to dominate the X-rays. Once this is the case, the
ratio $L_{\rm core}/L_{\rm x}$ will stay constant and reflect simply
the ratio intrinsic to the jet (assuming that the radio and X-ray
beaming parameters are the same).

To investigate if this behaviour is indeed observed we have included
for comparison in Fig. \ref{lrclx} also other types of radio-loud AGN,
selected as described in Section \ref{samples}. From Fig. \ref{lrclx}
we see that, as expected, the X-ray emission in SSRQ (filled
triangles), which are lobe-dominated sources and so are presumably
seen at larger angles, appears to be dominated by an unbeamed
component \citep{Kem93, Bak95b, Sie98}. Their ratios of radio core
emission to total X-ray luminosity $L_{\rm core}/L_{\rm x}$ increase
with $R$. On the other hand, in FSRQ (filled squares), which are
core-dominated sources and so are presumably seen at small angles, the
X-ray emission appears to be indeed dominated by the beamed jet. Their
ratios of $L_{\rm core}/L_{\rm x}$ are independent of $R$. In these
sources these ratios should then reflect the intrinsic jet SED, where
values of $\log L_{\rm core}/L_{\rm x} \ge 6$ and $<6$ are indicative
of X-rays dominated by inverse Compton and synchrotron emission,
respectively \citep{P96}.

However, so far, BL Lacs have been the only blazars known to have jets
with synchrotron X-rays. This is apparent in Fig. \ref{lrclx} where
these sources (open squares), almost all core-dominated, span a large
range in $L_{\rm core}/L_{\rm x}$ and reach relatively low values. On
the other hand, all but two of the ``classical'' FSRQ plotted here
have ratios of $L_{\rm core}/L_{\rm x} \ge 6$.

What is the origin of the X-ray emission in our newly discovered
``X-ray strong'' radio quasars? Their distribution in Fig. \ref{lrclx}
shows that a third (9/25 objects) fall in the range $R<1$ and follow
the behaviour exhibited by SSRQ. Therefore, the X-ray emission in
these sources is most likely {\it not} dominated by the jet, but comes
rather from the ambient hot gas and/or the the accretion disk. We have
performed correlation analysis for the following three cases, all SSRQ
(63 sources), SSRQ with detected X-ray emission (24 sources), and our
lobe-dominated sources. In the first case censored data was present
and we used the ASURV analysis package \citep{Iso86}. In all cases the
observed correlations are highly significant ($P>99.9\%$) with slopes
$0.79\pm0.11$, $0.88\pm0.09$, and $0.93\pm0.28$, respectively, not
significantly different from one (as expected for a roughly constant
ratio of $L_{\rm ext}/L_{\rm x}$). The best fits are shown as dashed
lines in Fig. \ref{lrclx}. At $R=1$, these correlations give the
average $\log L_{\rm core}/L_{\rm x}$ ratios intrinsic to the jets of
the corresponding population. Fig. \ref{lrclx} shows that this value
is for our sources $<6$, i.e., suggestive of synchrotron X-rays.

The majority of our sources (16/25 objects) have a core-dominated
radio morphology and, contrary to the majority of ``classical'' FSRQ,
their $L_{\rm core}/L_{\rm x}$ ratios extend well into the region
populated by high-energy peaked BL Lacs. In fact, almost all of our
core-dominated sources (14/16 objects) have ratios $\log L_{\rm
core}/L_{\rm x} \le 6$, indicative of synchrotron X-rays. The definite
proof of the synchrotron nature of their X-ray emission, however, will
require X-ray imaging and spectroscopy.

Dedicated X-ray observations exist or have been scheduled for 11/25
sources presented here. RGBJ1629$+$4008, the core-dominated radio
quasar with the lowest $L_{\rm core}/L_{\rm x}$ ratio in our sample,
has been observed with {\sl Beppo}SAX by \citet{P02} and found to have
X-rays dominated by synchrotron radiation with a peak frequency $\sim
2\times10^{16}$ Hz. A further four sources have been observed with
XMM-Newton in AO-3 (PI: Padovani) and six sources have been approved
for observation with {\sl Chandra} in Cycle 6 (PI: Landt). Out of
these 10 objects, 6 sources are core-dominated and have the lowest
$L_{\rm core}/L_{\rm x}$ ratios in our sample after RGBJ1629$+$4008.

\section{Summary and Conclusions}

A considerable fraction of flat-spectrum radio quasars discovered in
two recent blazar surveys (DXRBS and RGB) have multiwavelength
emission properties similar to those of BL Lacs with synchrotron
X-rays \citep{P03}. If it can be shown that these ``X-ray loud'' radio
quasars are indeed the strong-lined version of high-energy peaked BL
Lacs, it could mean that jet energetics are less dependent on emission
line luminosity than advocated so far \citep{Sam96, Fos98, Ghi98}.

In this paper we have presented deep VLA radio images obtained at 1.4
GHz in A and C configurations of 25 of these newly discovered blazars
(8 RGB and 17 DXRBS sources). We have compared the radio properties of
our sources with those of other classes of radio-loud AGN, namely,
``classical'' steep- and flat-spectrum radio quasars and BL Lacs. Our
main results can be summarized as follows.

(i) The radio morphologies of our sources are similar to those
generally observed for radio quasars and include prominent cores,
lobes and jet features. Their range of extended radio powers, however,
is similar to that of BL Lacs and extends down to the values typical
of FR I radio galaxies, reaching about three orders of magnitude lower
values than ``classical'' radio quasars. This result is partly due to
the fact that our sources are selected from surveys with relatively
low radio flux limits.

(ii) In our sample, 9/25 sources are lobe-dominated and in principle a
morphological classification of their extended radio structure is
possible. Out of these, 4 sources have FR II-like extended powers with
two having a clear and another two a tentative FR II-like
morphology. Another 5 sources have extended radio powers typical of
both FR Is and FR IIs. Out of these, 4 show a clear FR II-like
morphology, whereas one source requires further observations since its
extended emission is resolved out on the scales imaged.

(iii) As expected from unified schemes, we find for our sources and
the comparison samples of radio-loud AGN that the largest linear sizes
(LLS) decrease with increasing radio core dominance parameter
$R$. Fits based on the relativistic beaming model to the upper and
lower envelopes of the observed LLS$- R$ relation yield bulk Lorentz
factors of $\Gamma = 6-8$ and maximum and minimum intrinsic LLS values
of $\sim 2000$ and $\sim 20$ kpc, respectively.

(iv) The majority of our sources (16/25 objects) have a core-dominated
radio morphology and thus their total X-ray emission is expected to be
dominated by the jet. Out of these, 14/16 have ratios of radio core to
total X-ray luminosity $\log L_{\rm core}/L_{\rm x} \le 6$, indicative
of synchrotron X-rays \citep{P96}. The definite proof of the
synchrotron nature of their X-ray emission, however, will require
dedicated observations with {\sl Chandra} and XMM-Newton. 

We plan to continue our VLA observing program of ``X-ray loud'' radio
quasars in order to find in particular lobe-dominated radio quasars
with low extended powers. These sources could then help us clarify if
a large population of quasars hosted by FR Is exists.

\acknowledgments 

H.L. acknowledges financial support from the Deutsche Akademie der
Naturforscher Leopoldina grant number BMBF-LPD 9901/8-99. This
research has made use of the NASA/IPAC Extragalactic Database (NED)
which is operated by the Jet Propulsion Laboratory, California
Institute of Technology, under contract with the National Aeronautics
and Space Administration.

%\small
%\bibliography{references}

\begin{deluxetable}{lcrrccccc}
\tabletypesize{\footnotesize} 
\tablecaption{
\label{general} 
General Properties of the Sample and Log of Observations
} 
\tablewidth{0pt} 
\tablehead{ 
\colhead{Object Name} & \colhead{z} & \colhead{$f_{\rm NVSS}$} &
\colhead{$\alpha_{\rm r}$} & \colhead{$f_{\rm 1keV}$} &
\colhead{$\alpha_{\rm rx}$} & \colhead{observed} & \colhead{Array} &
\colhead{program} \\ 
&& \colhead{[mJy]} & & \colhead{[$\mu$Jy]} & & & \\
\colhead{(1)} & \colhead{(2)} & \colhead{(3)} & \colhead{(4)} &
\colhead{(5)} & \colhead{(6)} & \colhead{(7)} & \colhead{(8)} & \colhead{(9)}
}
\startdata
RGB \hspace{1.1mm}J0112$+$3818 & 0.333 &  57.5 & $ $0.09 & 0.096 & 0.72 & 01/10/2001 & A & AP395 \\
                               &       &       &         &       &      & 05/01/2000 & C & AP395 \\ 
RGB \hspace{1.1mm}J0141$+$3923 & 0.080 & 116.5 & $ $0.30 & 0.126 & 0.75 & 01/10/2001 & A & AP395 \\
                               &       &       &         &       &      & 05/01/2000 & C & AP395 \\ 
RGB \hspace{1.1mm}J0254$+$3931 & 0.291 & 230.4 & $-$0.31 & 0.344 & 0.76 & 01/10/2001 & A & AP395 \\
                               &       &       &         &       &      & 05/01/2000 & C & AP395 \\ 
RGB \hspace{1.1mm}J1629$+$4008 & 0.272 &   9.0 & $-$0.29 & 0.211 & 0.61 & 11/14/2000 & A & AP395 \\
                               &       &       &         &       &      & 05/01/2000 & C & AP395 \\ 
RGB \hspace{1.1mm}J2229$+$3057 & 0.322 & 150.4 & $ $0.15 & 0.203 & 0.74 & 01/10/2001 & A & AP395 \\ 
                               &       &       &         &       &      & 05/01/2000 & C & AP395 \\
RGB \hspace{1.1mm}J2256$+$2618 & 0.121 &  43.6 & $-$0.02 & 0.208 & 0.69 & 11/14/2000 & A & AP395 \\
                               &       &       &         &       &      & 05/01/2000 & C & AP395 \\ 
RGB \hspace{1.1mm}J2308$+$2008 & 0.250 & 188.4 & $ $0.12 & 0.151 & 0.77 & 01/10/2001 & A & AP395 \\
                               &       &       &         &       &      & 05/01/2000 & C & AP395 \\ 
RGB \hspace{1.1mm}J2318$+$3048 & 0.103 &  23.1 & $ $0.12 & 0.185 & 0.65 & 11/14/2000 & A & AP395 \\ 
                               &       &       &         &       &      & 05/01/2000 & C & AP395 \\
WGA J0106$-$1034               & 0.469 & 276.0 & $ $0.35 & 0.091 & 0.79 & 03/24/2002 & A & AL564 \\
                               &       &       &         &       &      & 09/12/2005 & C & AL652 \\
WGA J0110$-$1647               & 0.781 & 105.6 & $ $0.35 & 0.176 & 0.69 & 11/28/2004 & A & AP479 \\
WGA J0126$-$0500               & 0.411 &  58.4 & $ $0.21 & 0.076 & 0.74 & 11/07/2004 & A & AL627 \\
WGA J0227$-$0847               & 2.228 &  65.3 & $-$0.34 & 0.051 & 0.70 & 11/28/2004 & A & AP479 \\
WGA J0259$+$1926               & 0.544 & 163.7 & $ $0.16 & 0.089 & 0.75 & 03/24/2002 & A & AL564 \\
WGA J0304$+$0002               & 0.563 & 123.8 & $ $0.40 & 0.121 & 0.74 & 03/24/2002 & A & AL564 \\
                               &       &       &         &       &      & 09/12/2005 & C & AL652 \\
WGA J0435$-$0811               & 0.791 &  50.1 & $-$0.27 & 0.028 & 0.75 & 02/09/2002 & A & AL564 \\
WGA J0447$-$0322               & 0.774 &  86.7 & $ $0.47 & 0.348 & 0.66 & 02/09/2002 & A & AL564 \\
                               &       &       &         &       &      & 10/18/2002 & C & AL578 \\
WGA J0544$-$2241               & 1.537 & 133.2 & $-$0.44 & 0.085 & 0.74 & 11/02/2004 & A & AL627 \\
                               &       &       &         &       &      & 10/03/2002 & CnB & AL578 \\
WGA J1026$+$6746               & 1.181 & 234.0 & $ $0.49 & 0.069 & 0.76 & 11/28/2004 & A & AP479 \\
WGA J1457$-$2818               & 1.999 & 225.4 & $ $0.40 & 0.019 & 0.74 & 12/04/2004 & A & AL627 \\
WGA J2239$-$0631               & 0.264 & 117.2 & $ $0.54 & 0.259 & 0.69 & 09/30/2004 & A & AL627 \\
                               &       &       &         &       &      & 09/12/2005 & C & AL652 \\
WGA J2320$+$0032               & 1.894 &  82.0 & $ $0.45 & 0.036 & 0.76 & 09/30/2004 & A & AL627 \\
WGA J2322$+$2114               & 0.707 & 151.6 & $ $0.31 & 0.064 & 0.77 & 09/29/2004 & A & AL627 \\
WGA J2347$+$0852               & 0.292 & 147.7 & $ $0.58 & 0.176 & 0.71 & 01/10/2001 & A & AP395 \\
                               &       &       &         &       &      & 05/01/2000 & C & AP395 \\
PKS 0256$-$005                 & 1.995 & 228.2 & $-$0.59 & 0.102 & 0.78 & 11/28/2004 & A & AP479 \\  
0959$+$68W1                    & 0.773 & 101.9 & $ $0.41 & 0.114 & 0.75 & 11/28/2004 & A & AP479 \\
\enddata
\end{deluxetable}

\begin{deluxetable}{lccrcrc}
\tabletypesize{\small}
\tablecaption{
\label{map} 
Map Parameters
}
\tablewidth{0pt}
\tablehead{
\colhead{Object Name} & \colhead{Array} & \colhead{beam} &
\colhead{PA} & \colhead{rms} & \colhead{peak} & \colhead{Fig. 1} \\
 & & \colhead{[arcsec]} & \colhead{[deg]} & \colhead{[mJy/} & 
\colhead{[mJy/} & \\
 & & & & \colhead{beam]} & \colhead{beam]} \\
\colhead{(1)} & \colhead{(2)} & \colhead{(3)} & \colhead{(4)} & 
\colhead{(5)} & \colhead{(6)} & \colhead{(7)}
}
\startdata
RGB \hspace{1.1mm}J0112$+$3818 & A     &  1.4$\times$ 1.3 & $-$60 & 0.04 &  32.5 & (a) \\
RGB \hspace{1.1mm}J0141$+$3923 & A     &  1.4$\times$ 1.3 & $-$52 & 0.12 & 104.5 & (b) \\
RGB \hspace{1.1mm}J0254$+$3931 & A     &  1.4$\times$ 1.3 & $ $73 & 0.08 & 160.2 & (c) \\
RGB \hspace{1.1mm}J0254$+$3931 & A$+$C &  6.0$\times$ 6.0 & $ $85 & 0.07 & 167.5 & (d) \\
RGB \hspace{1.1mm}J1629$+$4008 & A     &  2.0$\times$ 1.3 & $ $80 & 0.04 &   8.6 & (e) \\
RGB \hspace{1.1mm}J2229$+$3057 & A     &  1.4$\times$ 1.3 & $-$84 & 0.04 &  44.2 & (f) \\
RGB \hspace{1.1mm}J2229$+$3057 & A$+$C &  7.4$\times$ 6.5 & $ $49 & 0.04 &  44.2 & (g) \\
RGB \hspace{1.1mm}J2256$+$2618 & C     & 15.3$\times$14.5 & $-$84 & 0.06 &  24.7 & (h) \\
RGB \hspace{1.1mm}J2308$+$2008 & A     &  1.6$\times$ 1.4 & $ $66 & 0.06 & 218.0 & (i) \\
RGB \hspace{1.1mm}J2318$+$3048 & C     & 15.0$\times$14.4 & $-$90 & 0.06 &  22.2 & (j) \\
WGA J0106$-$1034               & A     &  2.0$\times$ 1.3 & $ $16 & 0.03 & 105.5 & (k) \\
WGA J0110$-$1647               & A     &  2.0$\times$ 1.3 & $ $10 & 0.10 &  72.0 & (l) \\
WGA J0126$-$0500               & A     &  1.7$\times$ 1.2 & $ $12 & 0.13 &  14.6 & (m) \\
WGA J0227$-$0847               & A     &  1.7$\times$ 1.3 & $-$ 7 & 0.10 &  76.7 & (n) \\
WGA J0259$+$1926               & A     &  1.4$\times$ 1.3 & $ $27 & 0.05 & 133.1 & (o) \\
WGA J0304$+$0002               & A     &  1.6$\times$ 1.4 & $ $15 & 0.03 &   8.4 & (p) \\
WGA J0304$+$0002               & A$+$C &  6.5$\times$ 5.4 & $ $51 & 0.04 &  29.1 & (q) \\
WGA J0435$-$0811               & A     &  1.8$\times$ 1.4 & $ $ 3 & 0.04 &  30.3 & (r) \\
WGA J0447$-$0322               & A     &  1.6$\times$ 1.4 & $ $ 4 & 0.09 &  68.0 & (s) \\
WGA J0544$-$2241               & A     &  2.7$\times$ 1.3 & $-$19 & 0.06 & 149.6 & (t) \\
WGA J1026$+$6746               & A     &  2.7$\times$ 1.3 & $-$66 & 0.12 &  84.7 & (u) \\
WGA J1457$-$2818               & A     &  2.7$\times$ 1.3 & $-$ 4 & 0.16 & 190.4 & (v) \\
WGA J2239$-$0631               & C     & 17.8$\times$15.9 & $ $ 3 & 0.07 &  52.6 & (w) \\
WGA J2320$+$0032               & A     &  1.5$\times$ 1.4 & $-$14 & 0.15 &  65.6 & (x) \\
WGA J2322$+$2114               & A     &  1.4$\times$ 1.2 & $-$53 & 0.04 &  40.4 & (y) \\
WGA J2347$+$0852               & A     &  1.5$\times$ 1.4 & $ $53 & 0.06 &  19.4 & (z) \\
WGA J2347$+$0852               & A$+$C &  8.5$\times$ 8.4 & $-$38 & 0.08 &  52.3 & (aa) \\
PKS 0256$-$005                 & A     &  1.6$\times$ 1.4 & $-$14 & 0.05 & 252.1 & (ab)\\
0959+68W1                      & A     &  2.6$\times$ 1.3 & $-$70 & 0.06 &  80.1 & (ac)\\
\enddata
\end{deluxetable}

\begin{deluxetable}{lllrrcrrcr}
\tabletypesize{\small}
\rotate
\tablewidth{22cm}
\tablecaption{
\label{radioprop} 
Observed 20 Centimeter Radio Properties of the Sample
}
\tablewidth{0pt}
\tablehead{
\colhead{Object Name} & \colhead{R.A.(J2000)} & \colhead{Decl.(J2000)} & \colhead{$f_{\rm core}$} & 
\colhead{$f_{\rm ext}$} & \colhead{$\log L_{\rm core}$} & \colhead{$\log L_{\rm ext}$} & 
\colhead{$\log R$} & \colhead{$f_{\rm NVSS}/$} & \colhead{LAS} \\
                      &                       &                        & \colhead{[mJy]}          & 
\colhead{[mJy]}         & \colhead{[W/Hz]}              & \colhead{[W/Hz]}             & 
                   & \colhead{$f_{\rm total}$} & \colhead{[arcsec]} \\
\colhead{(1)}         & \colhead{(2)}         & \colhead{(3)}          & \colhead{(4)}            & 
\colhead{(5)}           & \colhead{(6)}                 & \colhead{(7)}                & 
\colhead{(8)}      & \colhead{(9)}            & \colhead{(10)}
}
\startdata
RGB \hspace{1.1mm}J0112$+$3818 & 01 12 18.049 & $+$38 18 56.90 &  32.5 &   5.8$\pm$0.4 & 25.20 &    24.55 &    0.65 & 1.50 &   20.6 \\
RGB \hspace{1.1mm}J0141$+$3923 & 01 41 57.756 & $+$39 23 29.10 & 104.5 &        $<$0.2 & 24.46 & $<$21.77 & $>$2.69 & 1.11 & $<$1.4 \\
RGB \hspace{1.1mm}J0254$+$3931 & 02 54 42.629 & $+$39 31 34.75 & 160.2 &  74.4$\pm$1.1 & 25.77 &    25.53 &    0.24 & 0.98 &   34.4 \\
RGB \hspace{1.1mm}J1629$+$4008 & 16 29 01.336 & $+$40 08 00.10 &   8.6 &        $<$0.2 & 24.44 & $<$22.89 & $>$1.55 & 1.05 & $<$2.0 \\
RGB \hspace{1.1mm}J2229$+$3057 & 22 29 34.151 & $+$30 57 12.10 &  44.2 & 116.5$\pm$0.4 & 25.30 &    25.82 & $-$0.52 & 0.94 &   90.6 \\
RGB \hspace{1.1mm}J2256$+$2618 & 22 56 39.163 & $+$26 18 43.55 &  21.4 &  12.6$\pm$0.4 & 24.13 &    23.94 &    0.19 & 1.28 &   65.9 \\
RGB \hspace{1.1mm}J2308$+$2008 & 23 08 11.630 & $+$20 08 42.05 & 218.0 &   9.7$\pm$0.3 & 25.77 &    24.50 &    1.27 & 0.83 &    7.1 \\
RGB \hspace{1.1mm}J2318$+$3048 & 23 18 36.905 & $+$30 48 37.00 &  17.5 &  12.1$\pm$0.4 & 23.90 &    23.78 &    0.12 & 0.78 &   41.5 \\
WGA J0106$-$1034               & 01 06 44.158 & $-$10 34 10.55 & 105.5 & 141.9$\pm$0.3 & 26.02 &    26.28 & $-$0.26 & 1.12 &    9.0 \\
WGA J0110$-$1647               & 01 10 35.516 & $-$16 48 27.80 &  72.0 &  27.0$\pm$0.8 & 26.31 &    26.09 &    0.22 & 1.07 &   31.8 \\
WGA J0126$-$0500               & 01 26 15.090 & $-$05 01 22.10 &  14.6 &  27.8$\pm$2.6 & 25.04 &    25.44 & $-$0.40 & 1.38 &   40.7 \\
WGA J0227$-$0847               & 02 27 32.070 & $-$08 48 12.20 &  76.7 &        $<$0.2 & 27.36 & $<$26.18 & $>$1.18 & 0.85 & $<$1.3 \\
WGA J0259$+$1926               & 02 59 29.650 & $+$19 25 44.35 & 133.1 &  11.8$\pm$0.3 & 26.25 &    25.35 &    0.90 & 1.13 &    8.1 \\
WGA J0304$+$0002               & 03 04 58.973 & $+$00 02 35.85 &   8.4 & 118.3$\pm$0.3 & 25.08 &    26.38 & $-$1.30 & 0.98 &   42.4 \\
WGA J0435$-$0811               & 04 35 08.353 & $-$08 11 02.75 &  30.3 &  19.8         & 25.95 &    25.97 & $-$0.02 & 1.65 &        \\
WGA J0447$-$0322               & 04 47 54.727 & $-$03 22 42.20 &  68.0 &        $<$0.1 & 26.28 & $<$23.65 & $>$2.63 & 1.28 & $<$1.6 \\
WGA J0544$-$2241               & 05 44 07.560 & $-$22 41 09.85 & 149.6 &        $<$0.1 & 27.27 & $<$24.42 & $>$2.85 & 0.89 & $<$2.7 \\
WGA J1026$+$6746               & 10 26 33.850 & $+$67 46 12.10 &  84.7 & 158.7$\pm$1.3 & 26.77 &    27.27 & $-$0.50 & 1.02 &   44.6 \\
WGA J1457$-$2818               & 14 57 44.599 & $-$28 19 21.60 & 190.4 &  30.0$\pm$0.7 & 27.64 &    27.22 &    0.42 & 1.02 &    6.3 \\
WGA J2239$-$0631               & 22 39 46.832 & $-$06 31 50.55 &  46.8 &  69.1$\pm$0.4 & 25.15 &    25.40 & $-$0.25 & 1.01 &   50.6 \\
WGA J2320$+$0032               & 23 20 37.983 & $+$00 31 39.70 &  65.6 &        $<$0.3 & 27.12 & $<$25.15 & $>$1.97 & 1.25 & $<$1.5 \\
WGA J2322$+$2114               & 23 22 02.607 & $+$21 13 56.45 &  40.4 &  86.9$\pm$1.1 & 25.97 &    26.49 & $-$0.52 & 1.19 &   55.9 \\
WGA J2347$+$0852               & 23 47 38.144 & $+$08 52 46.35 &   8.3 & 136.2$\pm$0.8 & 24.49 &    25.79 & $-$1.30 & 1.02 &   22.3 \\
PKS 0256$-$005                 & 02 59 28.510 & $-$00 20 00.00 & 252.2 &        $<$0.1 & 27.76 & $<$24.74 & $>$3.02 & 0.90 & $<$1.6 \\
0959$+$68W1                    & 10 03 06.760 & $+$68 13 16.80 &  80.1 &  26.8$\pm$0.7 & 26.35 &    26.07 &    0.28 & 0.95 &   27.6 \\
\enddata
\end{deluxetable}

\begin{figure*}
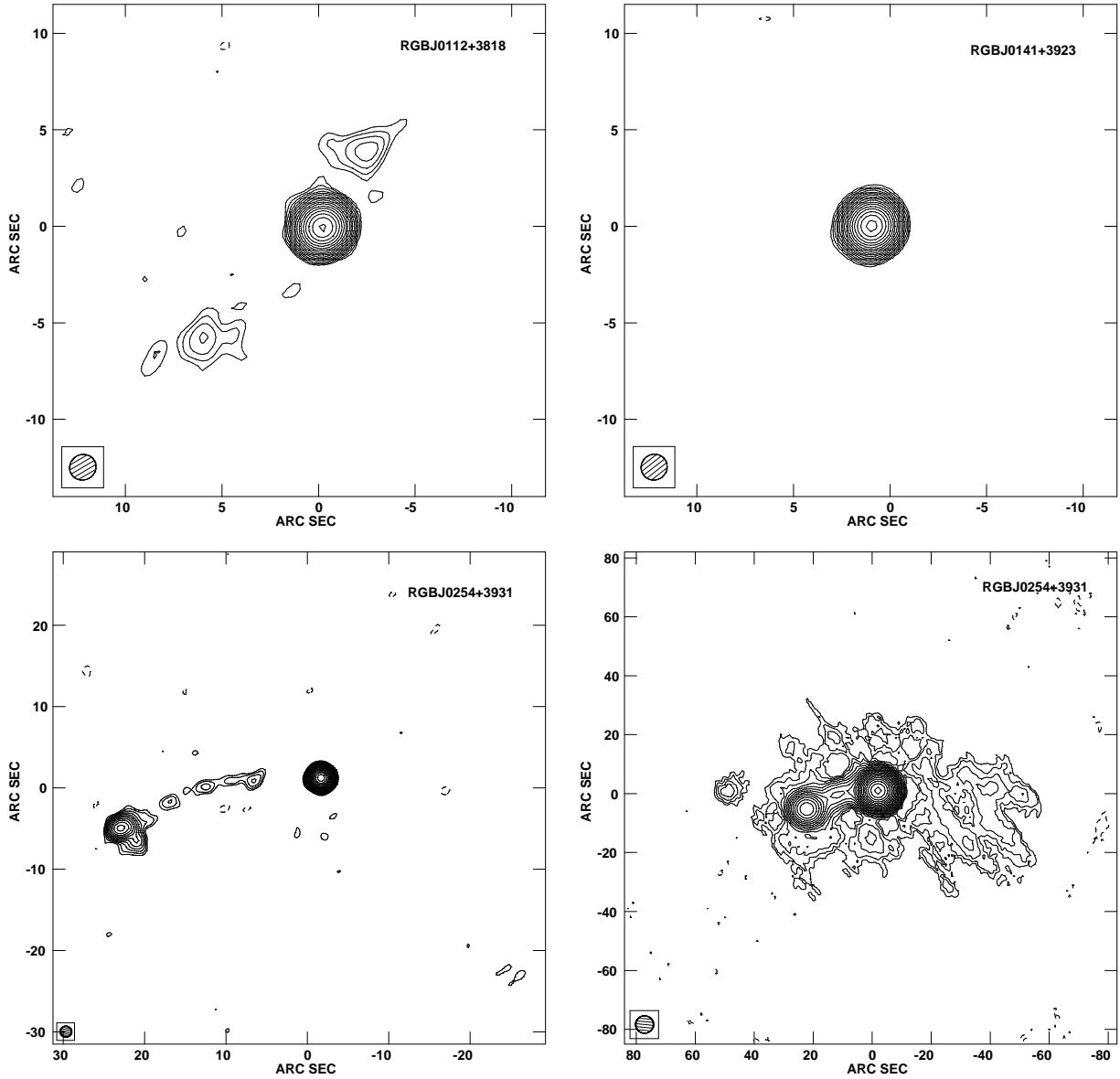

\centerline{
\includegraphics[scale=0.43]{f1a.eps}
\includegraphics[scale=0.43]{f1b.eps}
}
\centerline{
\includegraphics[scale=0.43]{f1c.eps}
\includegraphics[scale=0.43]{f1d.eps}
}
\caption{\small (a) RGB J0112$+$3818, VLA A. Image rms is 0.04
mJy/beam. Image peak is 32.5 mJy/beam. (b) RGB J0141$+$3923, VLA
A. Image rms is 0.12 mJy/beam. Image peak is 104.5 mJy/beam. (c) RGB
J0254$+$3931, VLA A. Image rms is 0.08 mJy/beam. Image peak is 160.2
mJy/beam. (d) RGB J0254$+$3931, VLA A$+$C. Image rms is 0.07
mJy/beam. Image peak is 167.5 mJy/beam. In all images contours start
at 3 times the rms and positive values are spaced by factors of
$\sqrt{2}$ up to the image peak.}
\end{figure*}

\setcounter{figure}{0}

\begin{figure*}
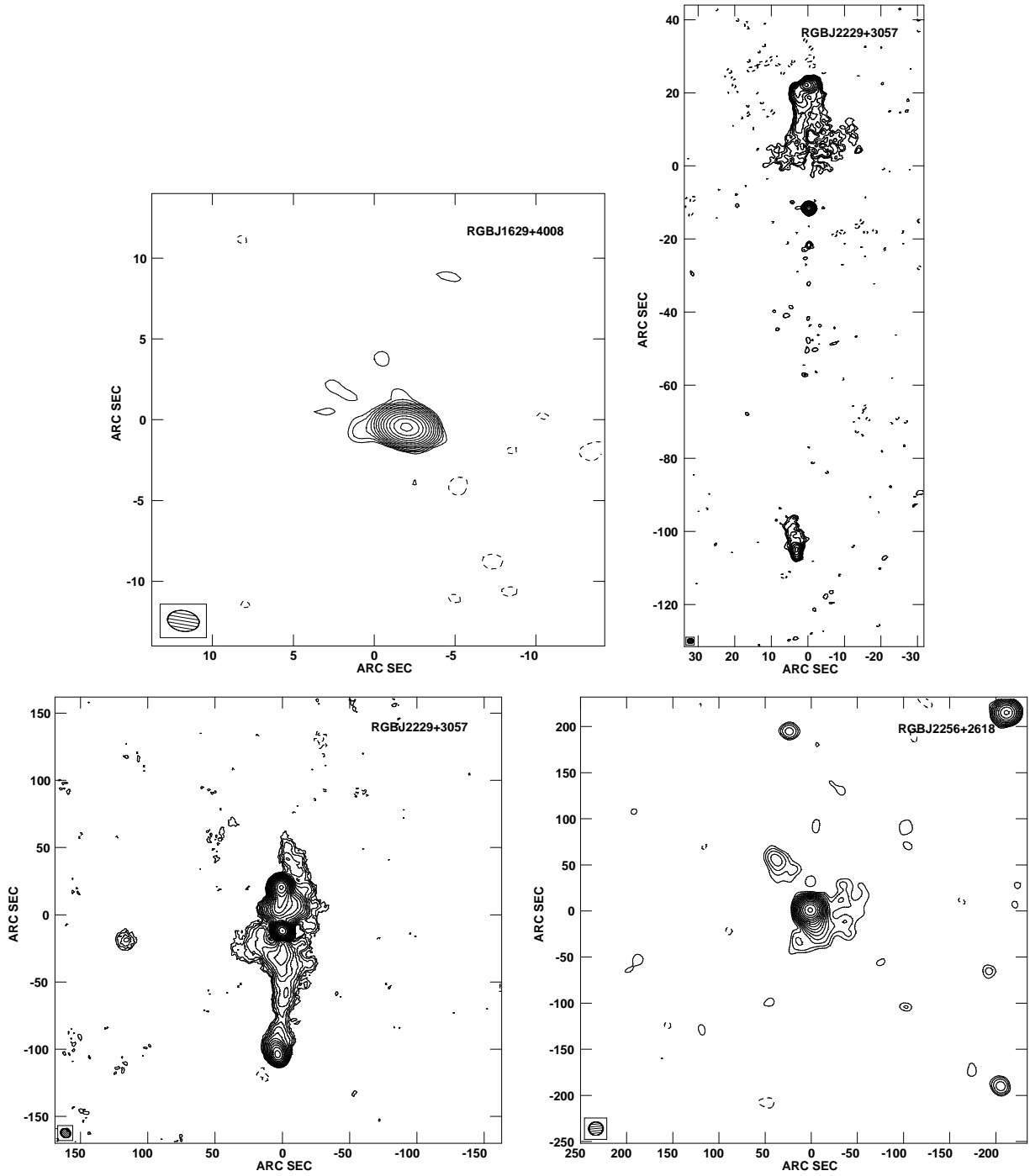

\centerline{
\includegraphics[scale=0.43]{f1e.eps}
\includegraphics[scale=0.43]{f1f.eps}
}
\centerline{
\includegraphics[scale=0.43]{f1g.eps}
\includegraphics[scale=0.43]{f1h.eps}
}
\caption{\small (e) RGB J1629$+$4008, VLA A. Image rms is 0.04
mJy/beam. Image peak is 8.6 mJy/beam. (f) RGB J2229$+$3057, VLA
A. Image rms is 0.04 mJy/beam. Image peak is 44.2 mJy/beam. (g) RGB
J2229$+$3057, VLA A$+$C. Image rms is 0.04 mJy/beam. Image peak is
44.2 mJy/beam. (h) RGB J2256$+$2618, VLA C. Image rms is 0.06
mJy/beam. Image peak is 24.7 mJy/beam. In all images contours start at
3 times the rms and positive values are spaced by factors of
$\sqrt{2}$ up to the image peak.}
\end{figure*}

\setcounter{figure}{0}

\begin{figure*}
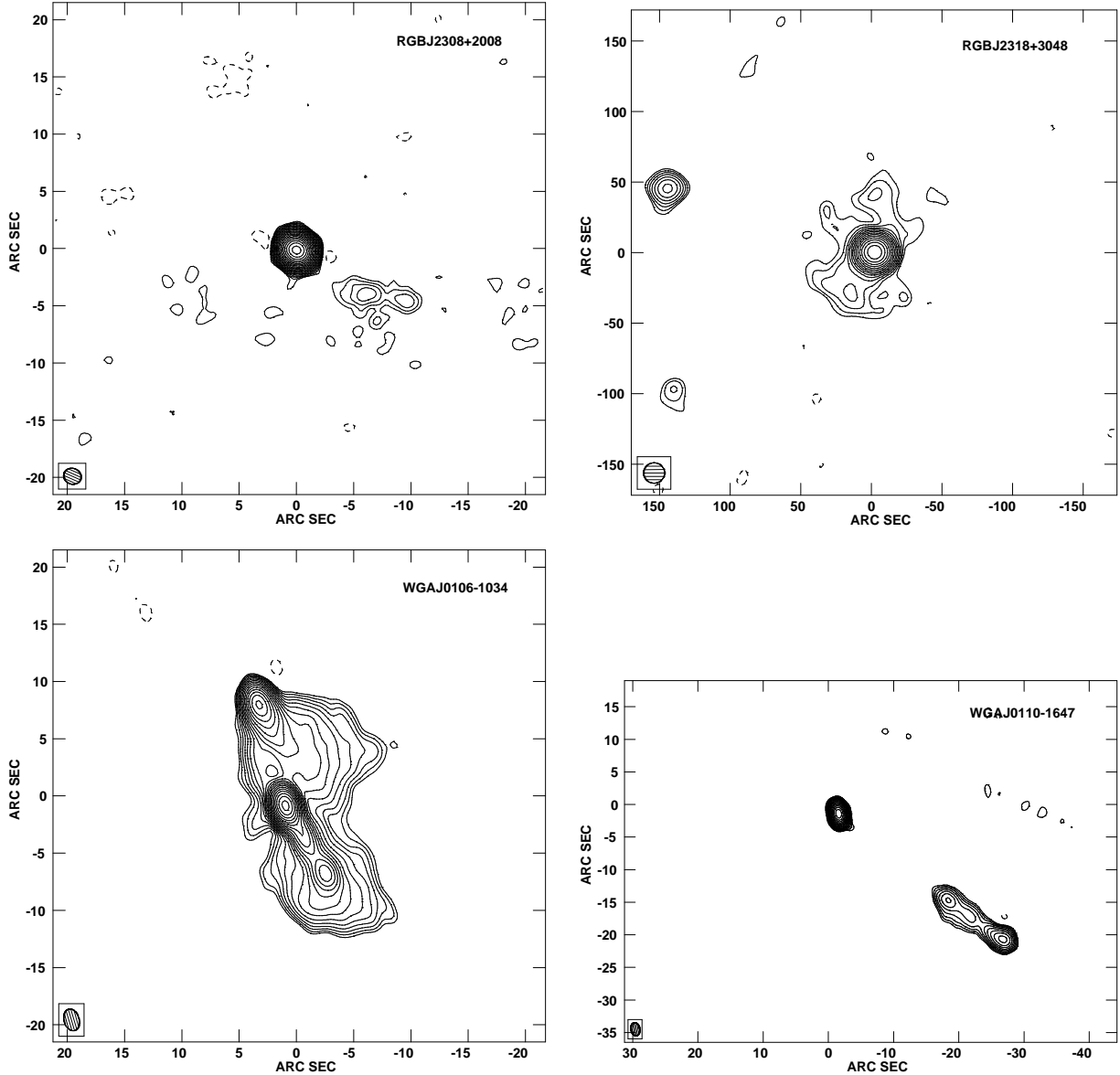

\centerline{
\includegraphics[scale=0.43]{f1i.eps}
\includegraphics[scale=0.43]{f1j.eps}
}
\centerline{
\includegraphics[scale=0.43]{f1k.eps}
\includegraphics[scale=0.43]{f1l.eps}
}
\caption{\small (i) RGB J2308$+$2008, VLA A. Image rms is 0.06
mJy/beam. Image peak is 218.0 mJy/beam. (j) RGB J2318$+$3048, VLA
C. Image rms is 0.06 mJy/beam. Image peak is 22.2 mJy/beam. (k) WGA
J0106$-$1034, VLA A. Image rms is 0.03 mJy/beam. Image peak is 105.5
mJy/beam. (l) WGA J0110$-$1647, VLA A. Image rms is 0.10
mJy/beam. Image peak is 72.0 mJy/beam. In all images contours start at
3 times the rms and positive values are spaced by factors of
$\sqrt{2}$ up to the image peak.}
\end{figure*}

\setcounter{figure}{0}

\begin{figure*}
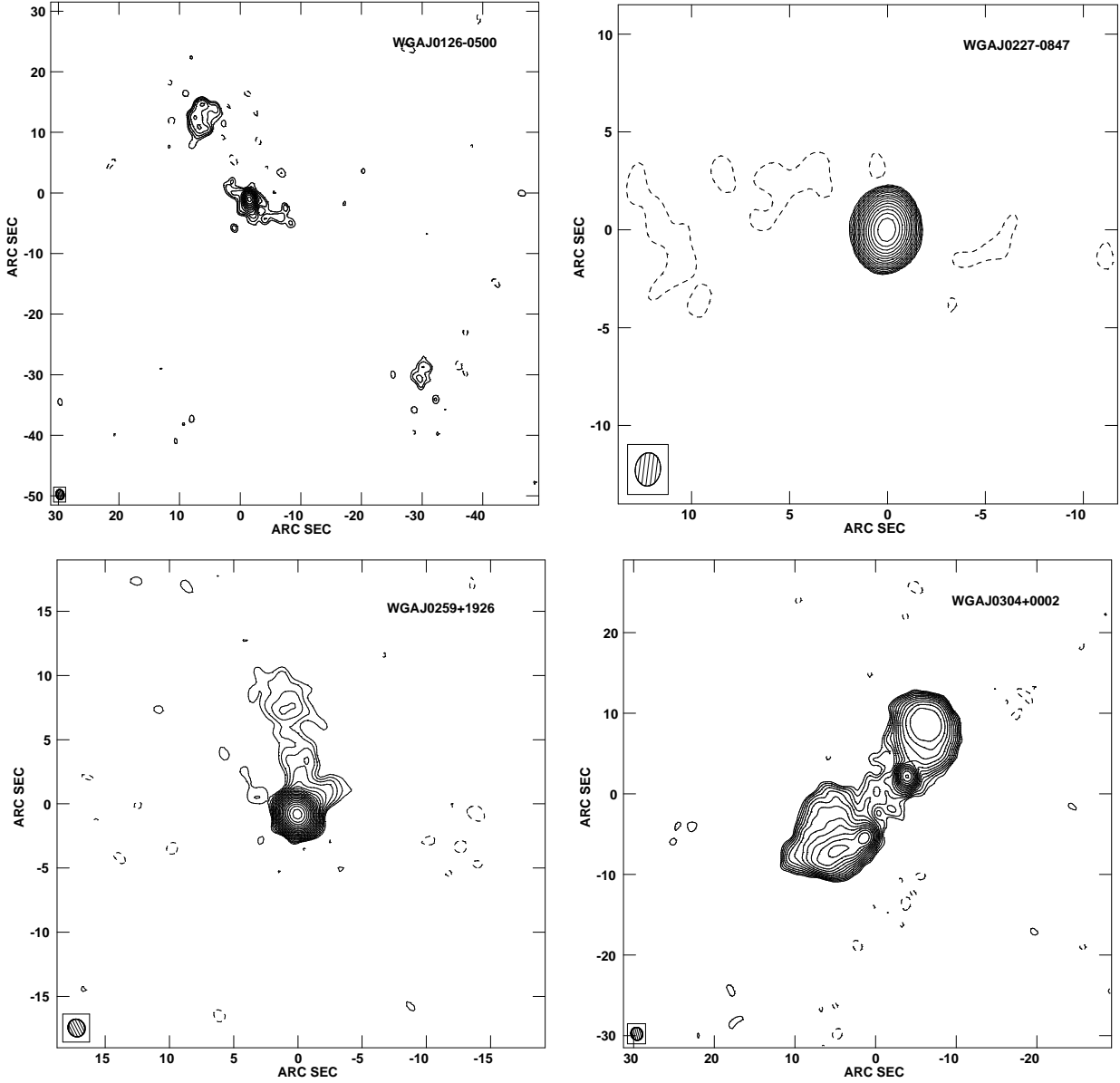

\centerline{
\includegraphics[scale=0.43]{f1m.eps}
\includegraphics[scale=0.44]{f1n.eps}
}
\centerline{
\includegraphics[scale=0.43]{f1o.eps}
\includegraphics[scale=0.43]{f1p.eps}
}
\caption{\small (m) WGA J0126$-$0500, VLA A. Image rms is 0.13
mJy/beam. Image peak is 14.6 mJy/beam. (n) WGA J0227$-$0847, VLA
A. Image rms is 0.10 mJy/beam. Image peak is 76.7 mJy/beam. (o) WGA
J0259$+$1926, VLA A. Image rms is 0.05 mJy/beam. Image peak is 133.1
mJy/beam. (p) WGA J0304$+$0002, VLA A. Image rms is 0.03
mJy/beam. Image peak is 8.4 mJy/beam. In all images contours start at
3 times the rms and positive values are spaced by factors of
$\sqrt{2}$ up to the image peak.}
\end{figure*}

\setcounter{figure}{0}

\begin{figure*}
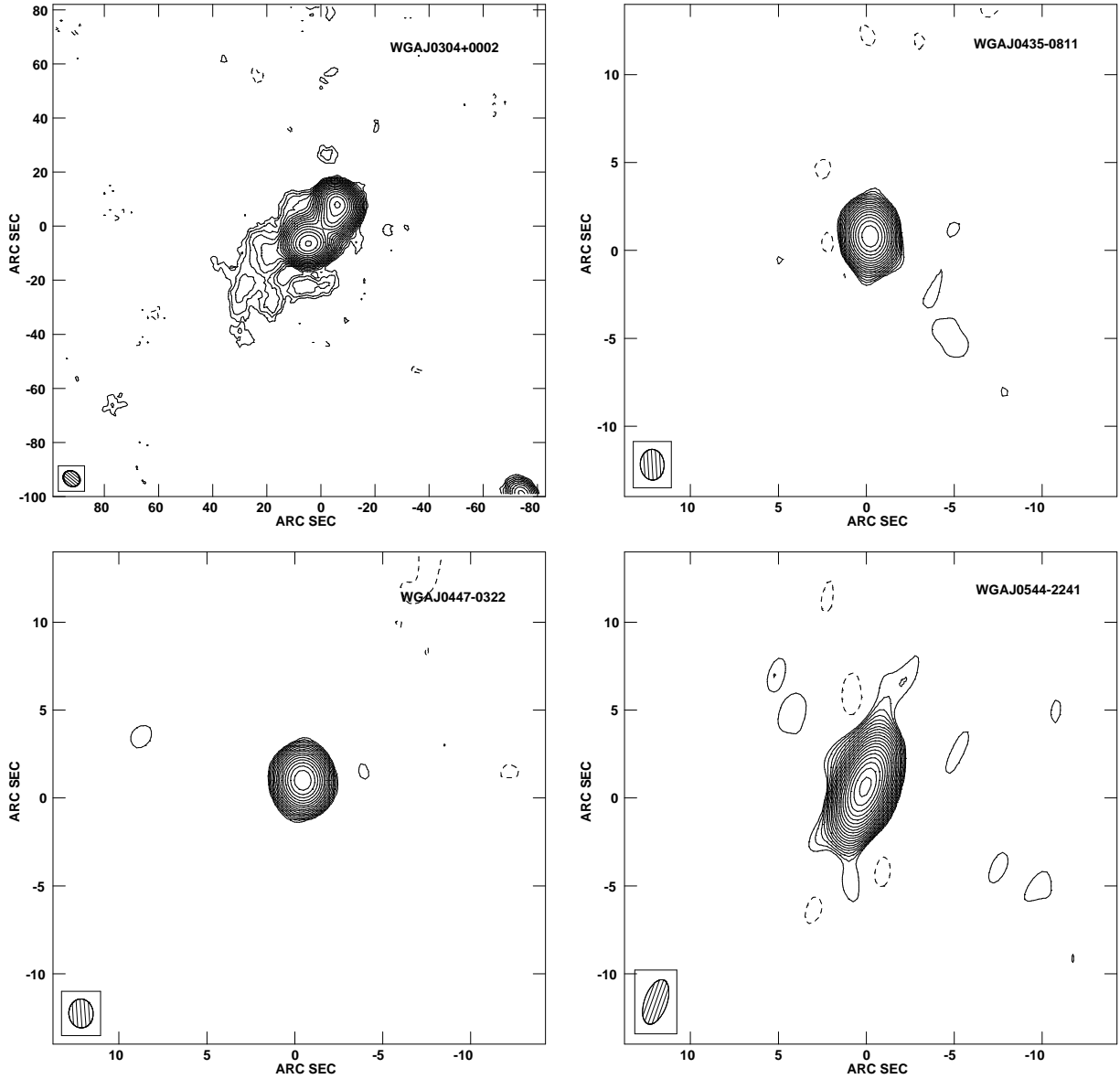

\centerline{
\includegraphics[scale=0.43]{f1q.eps}
\includegraphics[scale=0.43]{f1r.eps}
}
\centerline{
\includegraphics[scale=0.43]{f1s.eps}
\includegraphics[scale=0.43]{f1t.eps}
}
\caption{\small (q) WGA J0304$+$0002, VLA A$+$C. Image rms is 0.04
mJy/beam. Image peak is 29.1 mJy/beam. (r) WGA J0435$-$0811, VLA
A. Image rms is 0.04 mJy/beam. Image peak is 30.3 mJy/beam. (s) WGA
J0447$-$0322, VLA A. Image rms is 0.09 mJy/beam. Image peak is 68.0
mJy/beam. (t) WGA J0544$-$2241, VLA A. Image rms is 0.06
mJy/beam. Image peak is 149.6 mJy/beam. In all images contours start
at 3 times the rms and positive values are spaced by factors of
$\sqrt{2}$ up to the image peak.}
\end{figure*}

\setcounter{figure}{0}

\begin{figure*}
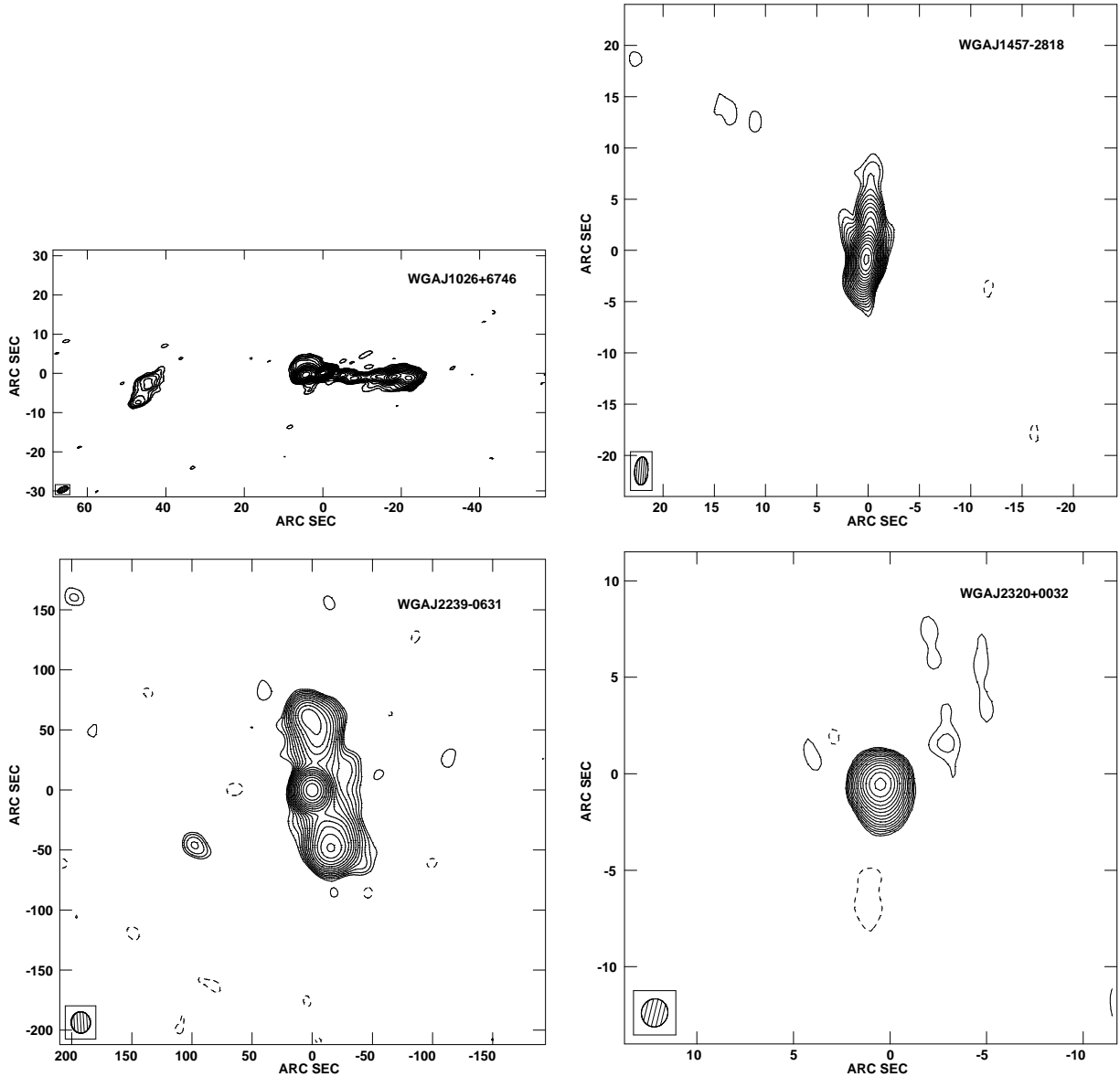

\centerline{
\includegraphics[scale=0.43]{f1u.eps}
\includegraphics[scale=0.43]{f1v.eps}
}
\centerline{
\includegraphics[scale=0.43]{f1w.eps}
\includegraphics[scale=0.43]{f1x.eps}
}
\caption{\small (u) WGA J1026$+$6746, VLA A. Image rms is 0.12
mJy/beam. Image peak is 84.7 mJy/beam. (v) WGA J1457$-$2818, VLA
A. Image rms is 0.16 mJy/beam. Image peak is 190.4 mJy/beam. (w) WGA
J2239$-$0631, VLA C. Image rms is 0.07 mJy/beam. Image peak is 52.6
mJy/beam. (x) WGA J2320$+$0032, VLA A. Image rms is 0.15
mJy/beam. Image peak is 65.6 mJy/beam. In all images contours start at
3 times the rms and positive values are spaced by factors of
$\sqrt{2}$ up to the image peak.}
\end{figure*}

\setcounter{figure}{0}

\begin{figure*}
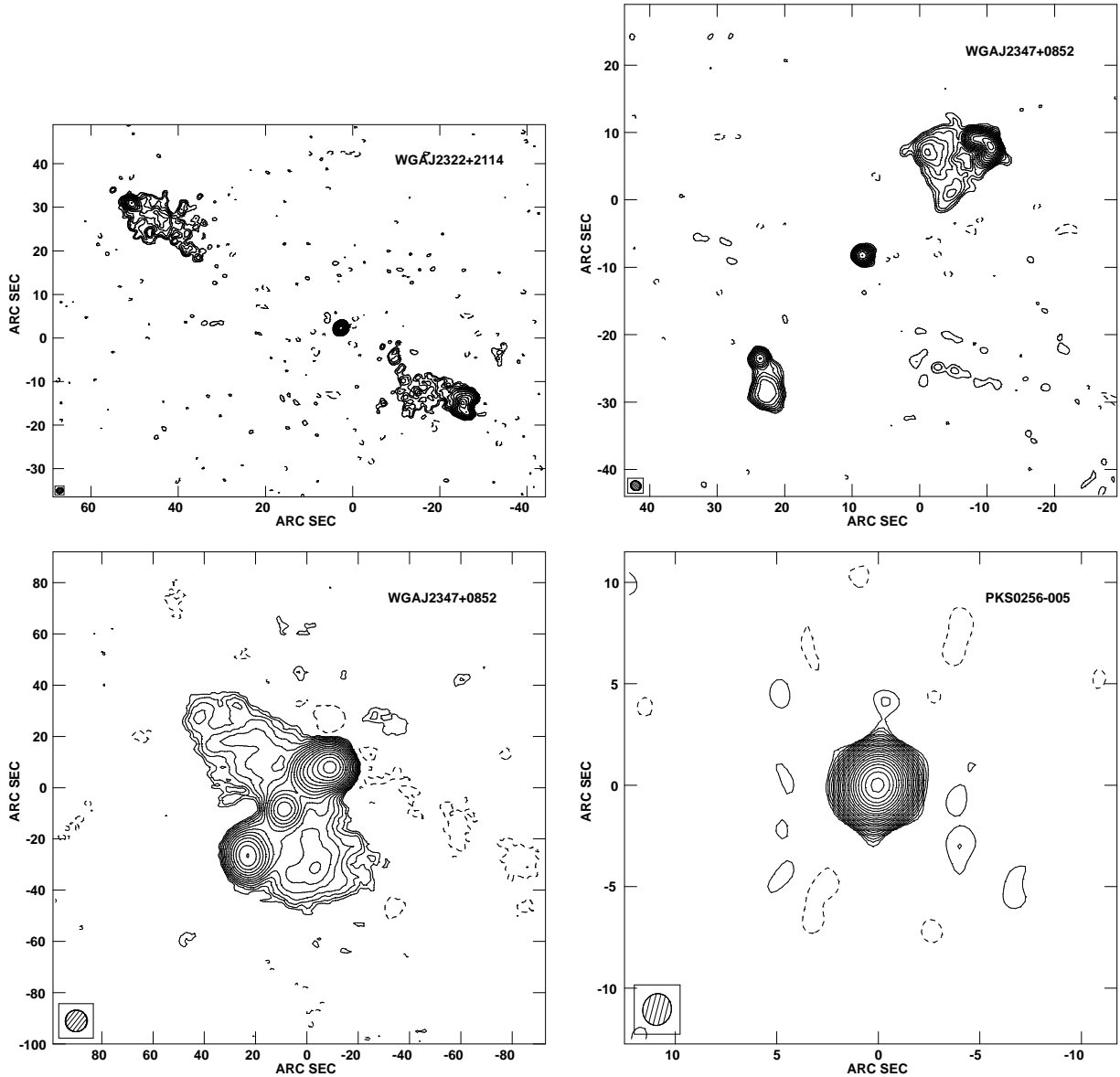

\centerline{
\includegraphics[scale=0.43]{f1y.eps}
\includegraphics[scale=0.43]{f1z.eps}
}
\centerline{
\includegraphics[scale=0.43]{f1aa.eps}
\includegraphics[scale=0.43]{f1ab.eps}
}
\caption{\small (y) WGA J2322$+$2114, VLA A. Image rms is 0.04
mJy/beam. Image peak is 40.4 mJy/beam. (z) WGA J2347$+$0852, VLA
A. Image rms is 0.06 mJy/beam. Image peak is 19.4 mJy/beam. (aa) WGA
J2347$+$0852, VLA A$+$C. Image rms is 0.08 mJy/beam. Image peak is
52.3 mJy/beam. (ab) PKS 0256$-$005, VLA A. Image rms is 0.05
mJy/beam. Image peak is 252.1 mJy/beam. In all images contours start
at 3 times the rms and positive values are spaced by factors of
$\sqrt{2}$ up to the image peak.}
\end{figure*}

\setcounter{figure}{0}

\begin{figure*}
\centerline{
\includegraphics[scale=0.43]{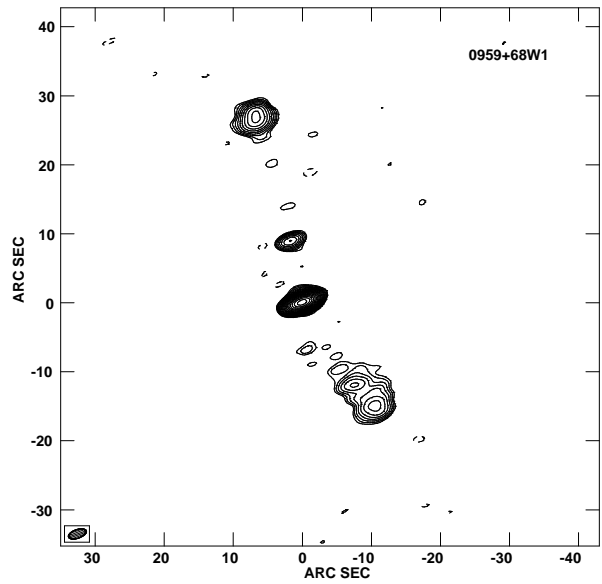}
}
\caption{\small (ac) 0959$+$68W1, VLA A. Image rms is 0.06
mJy/beam. Image peak is 80.1 mJy/beam. Contours start at 3 times the
rms and positive values are spaced by factors of $\sqrt{2}$ up to the
image peak.}
\end{figure*}

\clearpage

\begin{figure}
\plotone{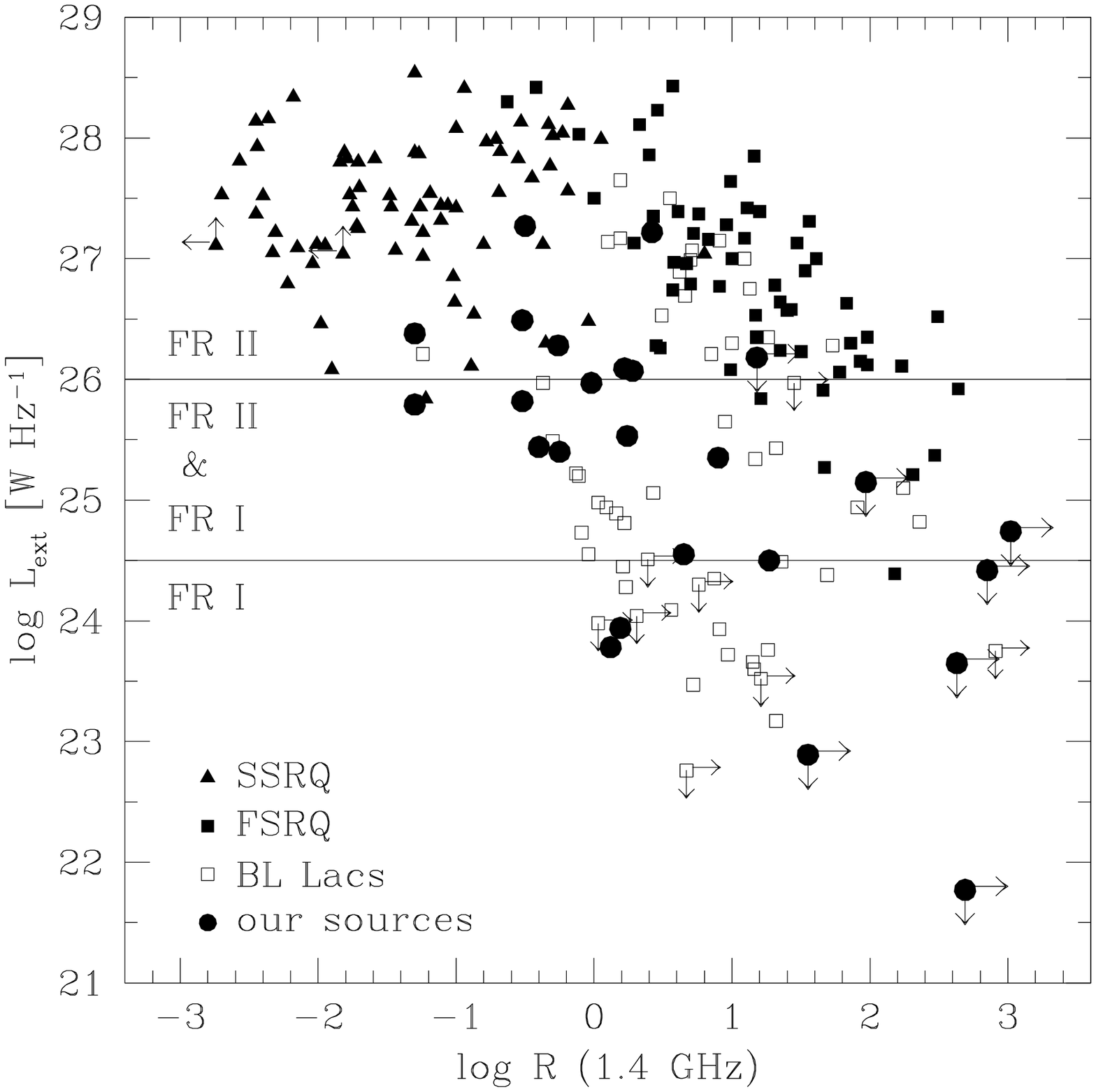}
\caption{
\label{logRlext}
The extended radio power versus the radio core dominance parameter $R$
at 1.4 GHz, where $R=L_{\rm core}/L_{\rm ext}$, with $L_{\rm core}$
and $L_{\rm ext}$ the radio core and extended luminosities,
respectively. Filled triangles and squares indicate steep-spectrum
(SSRQ) and flat-spectrum radio quasars (FSRQ), respectively, open
squares indicate BL Lacs. See text for details on the selected
samples. Large filled circles represent the sources in our
sample. Arrows indicate limits. The horizontal solid lines mark the
regions of extended radio powers typical of FR I and FR II radio
galaxies.}
\end{figure}

\begin{figure}
\plotone{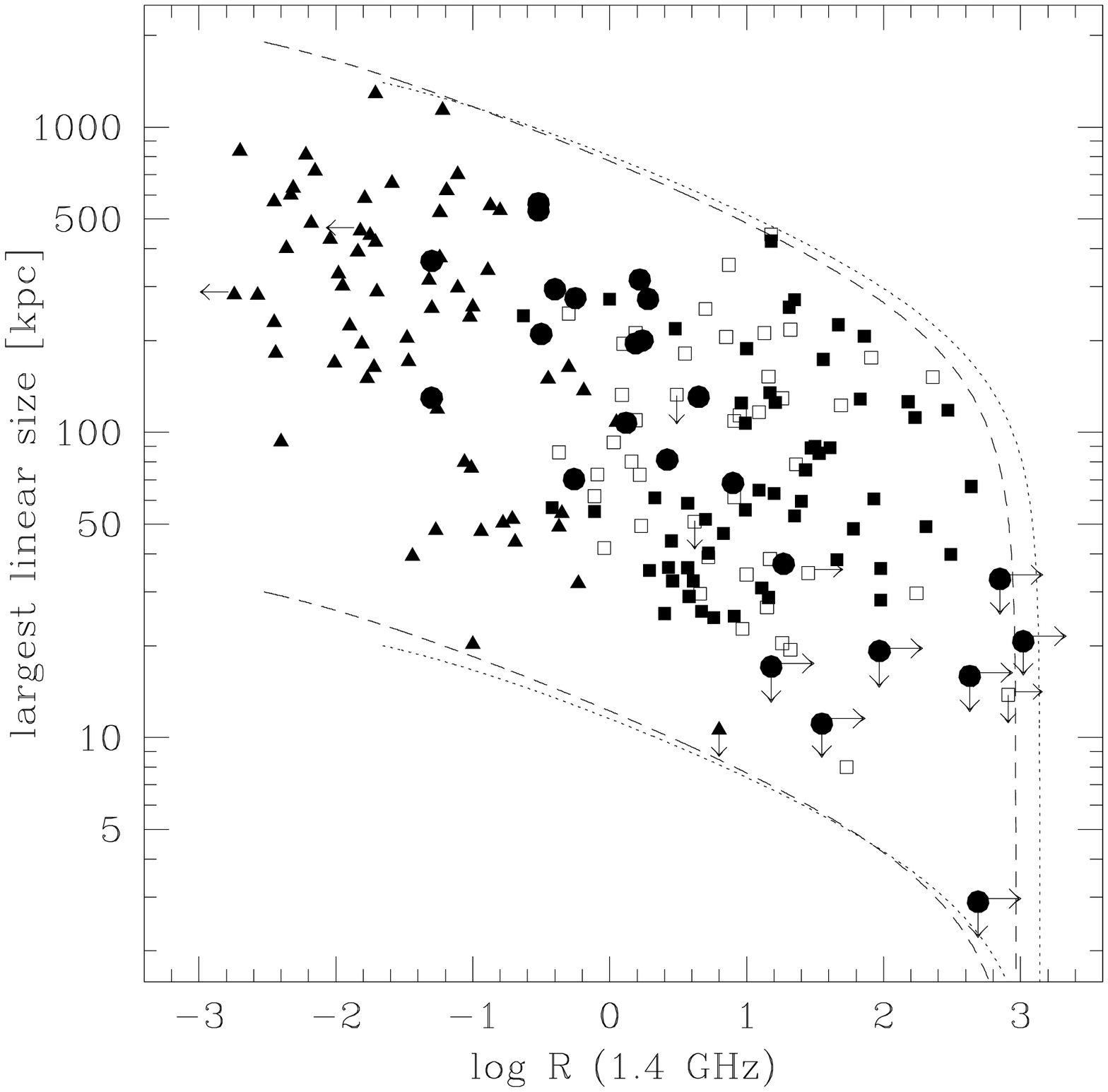}
\caption{
\label{logRLLS}
The largest linear size (LLS) versus the radio core dominance
parameter $R$ at 1.4 GHz. Symbols are as in Fig. \ref{logRlext}. The
dotted and dashed lines represent LLS $-$ $R$ relations predicted by
the relativistic beaming model for FR I and FR II radio galaxies,
respectively, assuming different intrinsic largest linear sizes. See
text for details.}
\end{figure}

\begin{figure}
\plotone{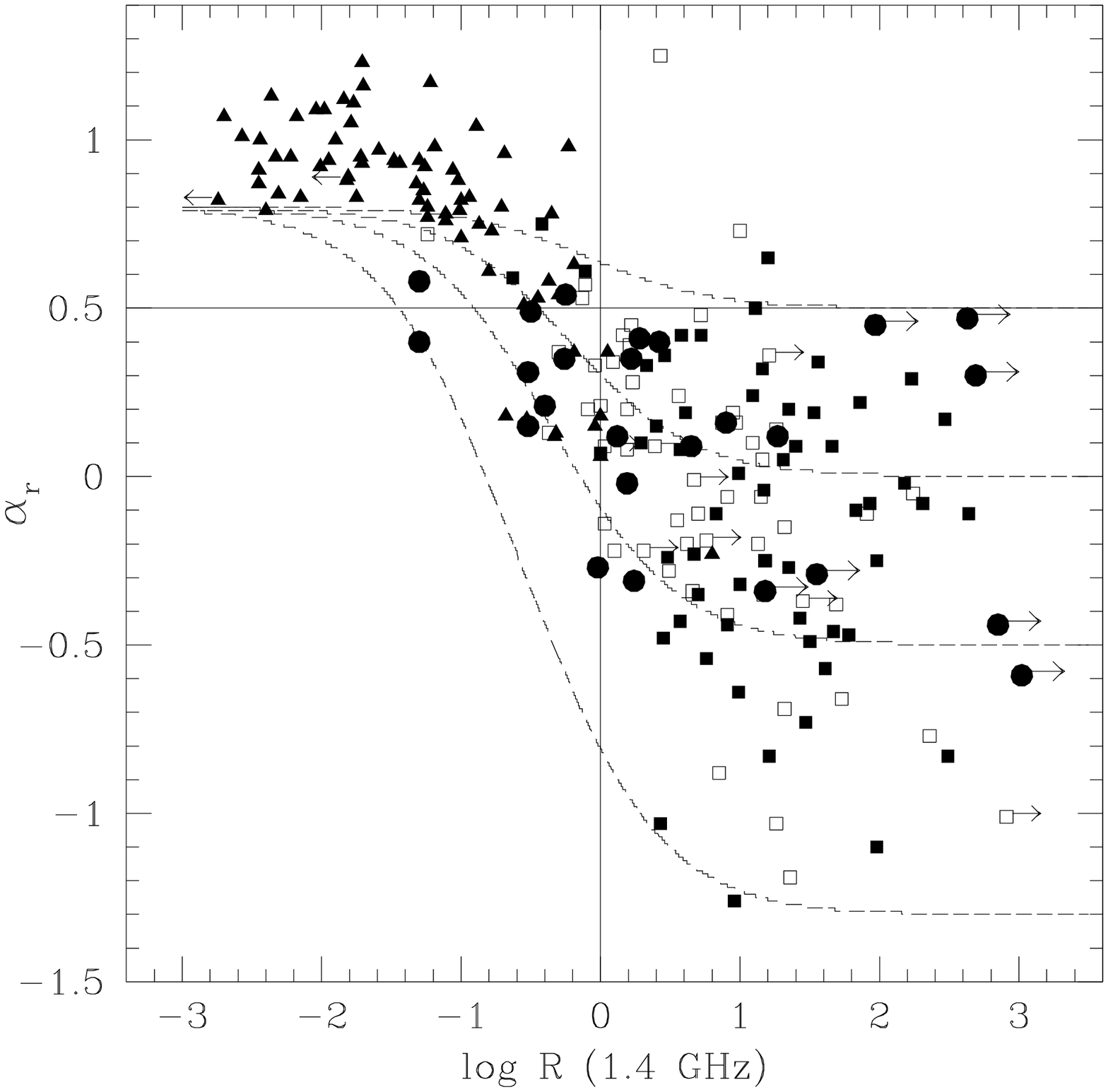}
\caption{
\label{logRalr}
The radio spectral index versus the radio core dominance parameter $R$
at 1.4 GHz. Symbols are as in Fig. \ref{logRlext}. The vertical solid
line separates lobe- (left of the line) from core-dominated sources
(right of the line). The horizontal solid line indicates the locus of
constant $\alpha_{\rm r}=0.5$. Radio-loud AGN below this line are
currently defined as blazars. The dashed lines represent $\alpha_{\rm
r} - R$ relations for different values of the core radio spectral
index of 0.5, 0, $-0.5$, and $-1.3$ (from top to bottom).}
\end{figure}

\begin{figure}
\plotone{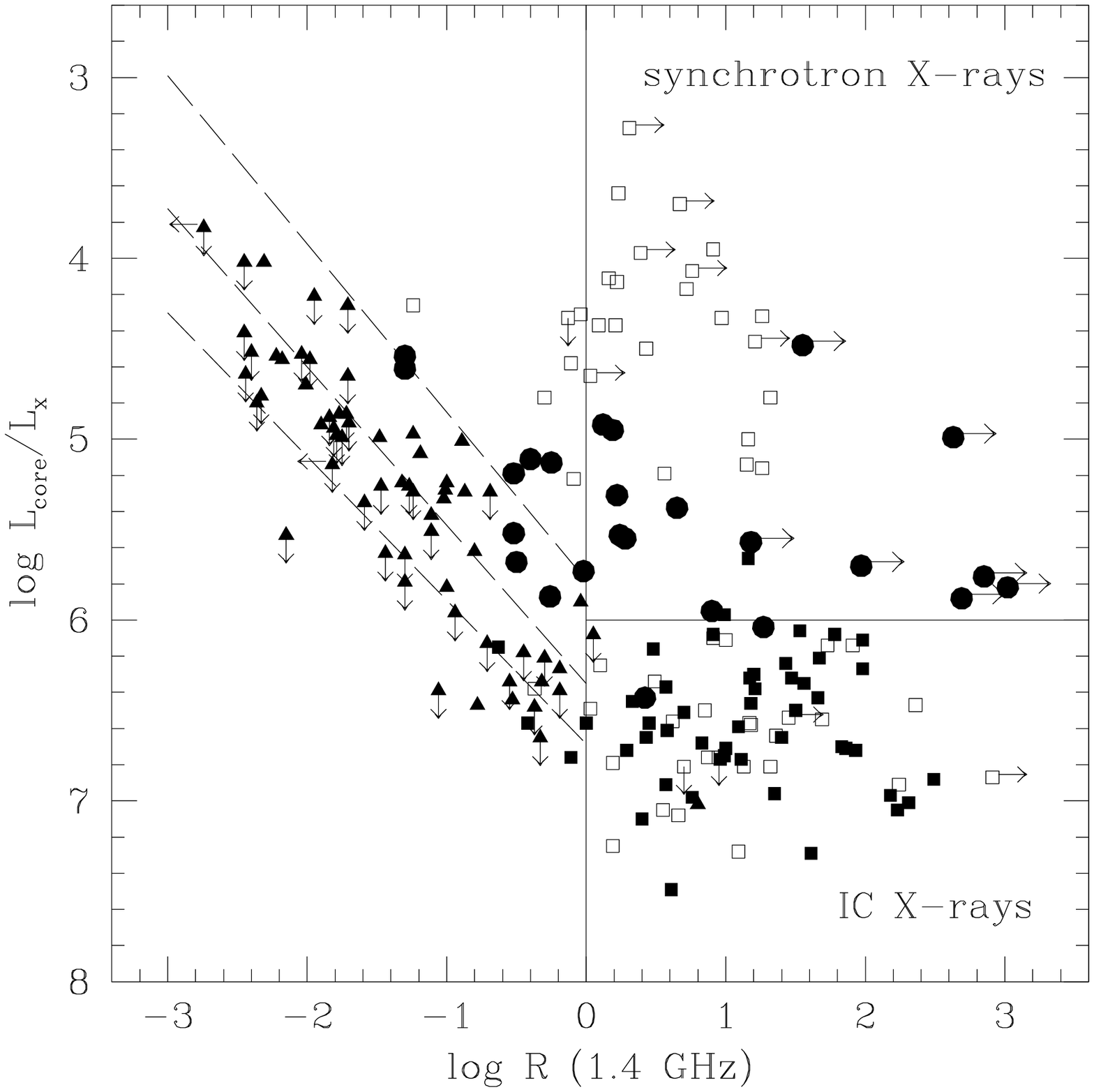}
\caption{\label{lrclx}
The ratio of radio core emission at 1.4 GHz to total X-ray luminosity
at 1 keV versus the radio core dominance parameter $R$ at 1.4
GHz. Symbols are as in Fig. \ref{logRlext}. The vertical solid line
separates lobe- (left of the line) from core-dominated sources (right
of the line). The horizontal solid line separates sources with jet
X-ray emission dominated by the synchrotron (above the line) and
inverse Compton process (below the line). The dashed lines represent
significant correlations for our lobe-dominated sources, SSRQ with
detected X-rays, and all SSRQ (from top to bottom).}
\end{figure}

\end{document}